\documentclass[aps,twocolumn,showpacs,nofootinbib,showkeys]{revtex4-1}

\usepackage{bm}
\usepackage{amsmath}
 \usepackage{amssymb}
 \usepackage{graphicx}
\usepackage{subfigure}
\usepackage{epsfig}
\usepackage{color}

\begin{document}

\title{Ordered Hexagonal Patterns via Notch-Delta Signaling}
\author{Eial Teomy}\thanks{Current Address: School of Mechanical Engineering, Tel Aviv University, Tel Aviv 69978, Israel}
\affiliation{Department of Physics, Bar-Ilan University, Ramat-Gan 52900, Israel}
\author{David A. Kessler }
\email{kessler@dave.ph.biu.ac.il}
\affiliation{Department of Physics, Bar-Ilan University, Ramat-Gan 52900, Israel}

\author{Herbert Levine }
\email{levine.herbert@northeastern.edu}
\affiliation{Dept of Physics, Northeastern Univ, Boston MA \\ and 
Center for Theoretical Biological Physics, Rice University, Houston, TX 77005, U.S.A.}
\date{\today}

\begin{abstract}
Many developmental processes in biology utilize Notch-Delta signaling to construct an ordered pattern of cellular differentiation. This signaling modality is based on nearest-neighbor contact, as opposed to the more familiar mechanism driven by the release of diffusible ligands. Here, exploiting this ``juxtacrine" property, we present an exact treatment of the pattern formation problem via a system of nine coupled ordinary differential equations. The possible patterns that are realized for realistic parameters can be analyzed by considering a co-dimension 2 pitchfork bifurcation of this system. This analysis explains the observed prevalence of hexagonal patterns with high Delta at their center, as opposed to those with central high Notch levels.  We show that outside this range of parameters, in particular for low cis-coupling, a novel kind of pattern is produced, where high Delta cells have high Notch as well.  It also suggests that the biological system is only weakly first order, so that  an additional mechanism is required to generate  the observed defect-free patterns. We construct a simple strategy for producing such defect-free patterns.
\end{abstract}


\maketitle
\section{Introduction}
Biological cells can exist in a number of distinct phenotypes, even with a fixed genome. These phenotypes arise via multi-stability of the underlying dynamical network controlling cell behavior and allow cells to take on differentiated roles in overall organism function. It is clear that developmental processes must ensure that these phenotypes arise in the right place and  time, i.e., ensure the emergence of functional phenotypic patterns.

A well-studied case of such a system is that of Notch-Delta signaling~\cite{notch-review}. Various cells  contain Notch transmembrane receptors~\cite{receptors} that couple to Notch ligands such as Delta or Jagged on both the same cell (cis-coupling) and neighboring cells (trans-coupling).  Because of the manner by which Notch and Delta inhibit each other (see below), their interaction typically leads to an alternating ``salt and pepper" structure. This type of patterning is seen in systems ranging from eyes~\cite{eyes} and ears~\cite{ears} to intestines~\cite{intestines} and livers~\cite{livers}. As a general rule, the high Delta cells are the most specialized ones (for example, the  photoreceptors~\cite{photoreceptors}) and are surrounded by less differentiated high Notch supporting cells. Parenthetically, changes in the transcriptional regulation  utilizing the Delta-alternative Jagged ligand may be crucial for the role of Notch in cancer metastasis~\cite{jagged,metastasis}, but here we focus solely on Delta and its interplay with Notch.
\begin{figure}[t]
\centering
\includegraphics[width=0.75\columnwidth]{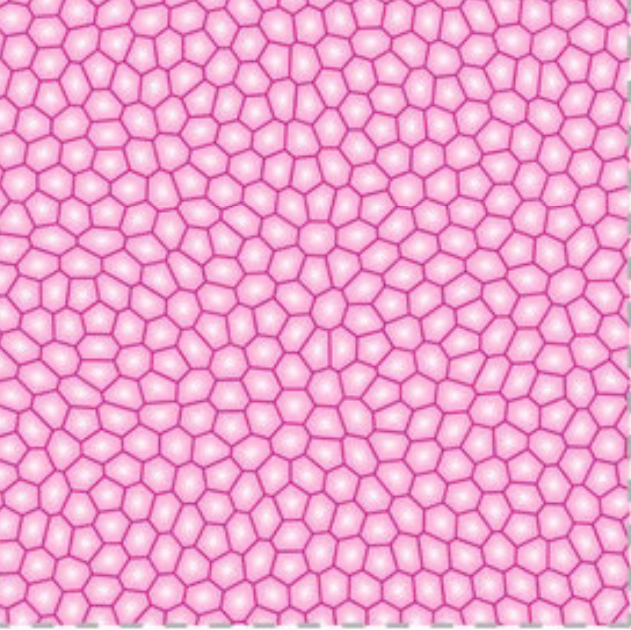}
\caption{Epithelial layer of MDCK cells; reproduced from \protect{\cite{photo}}}
\label{hex-layer}
\end{figure}

In this paper, we  study two aspects of the Notch-Delta system on 2d hexagonal arrays of cells. A motivation for this choice is that epithelial layers consist of polygonal cells that roughly form a hexagonal lattice, albeit with some size dispersion and some defects (see Fig. \ref{hex-layer}). It is worthwhile to first work out how pattern formation works in the more idealized case of the perfect lattice and then afterwards consider possible effects of the irregularities; we do note in passing that some effort has already been devoted to understanding the role of variations in cell size \cite{shaya}. First we focus on the existence and stability of hexagonal patterns in this geometry, which allows an exact re-writing of the (ordered) pattern-forming problem as a nine-dimensional dynamical system. We numerically construct the phase diagram of the system, showing the variety of patterns that emerge. Many features of this system can be understood by expanding about a co-dimension two pitchfork bifurcation. Others can be captured by examining various limiting cases of the parameters. The second aspect concerns mechanisms for ordered patterns to emerge from generic initial conditions. Here we identify a possible role for an initiating wave, similar to what has been seen in at least some biological realizations~\cite{wave}.

\begin{figure}[t]
\centering
\includegraphics[width=0.85\columnwidth]{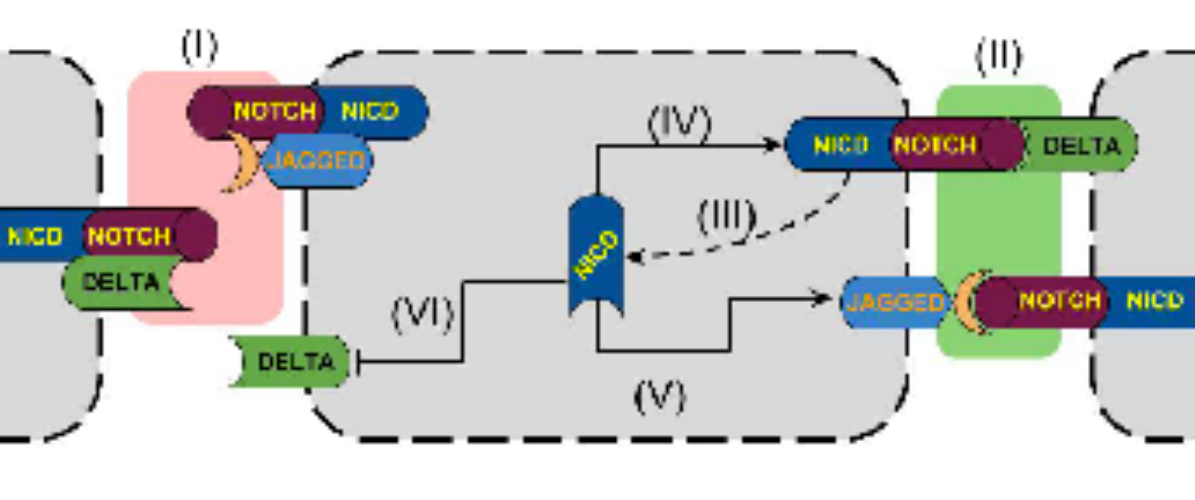}
\caption{Schematic showing Delta, Jagged and Notch-NICD complexes on the cell membrane (I). Binding of Notch on one cell to Delta on the other (II) leads to the freeing of the NICD, (III), which in turns leads to the enhancement of Notch (IV)  and Jagged (V) (which is irrelevant for our current concerns) and the suppression of Delta (VI).}
\label{juxtacrine}
\end{figure}

\begin{figure}[t]
\centering
\includegraphics[width=0.75\columnwidth]{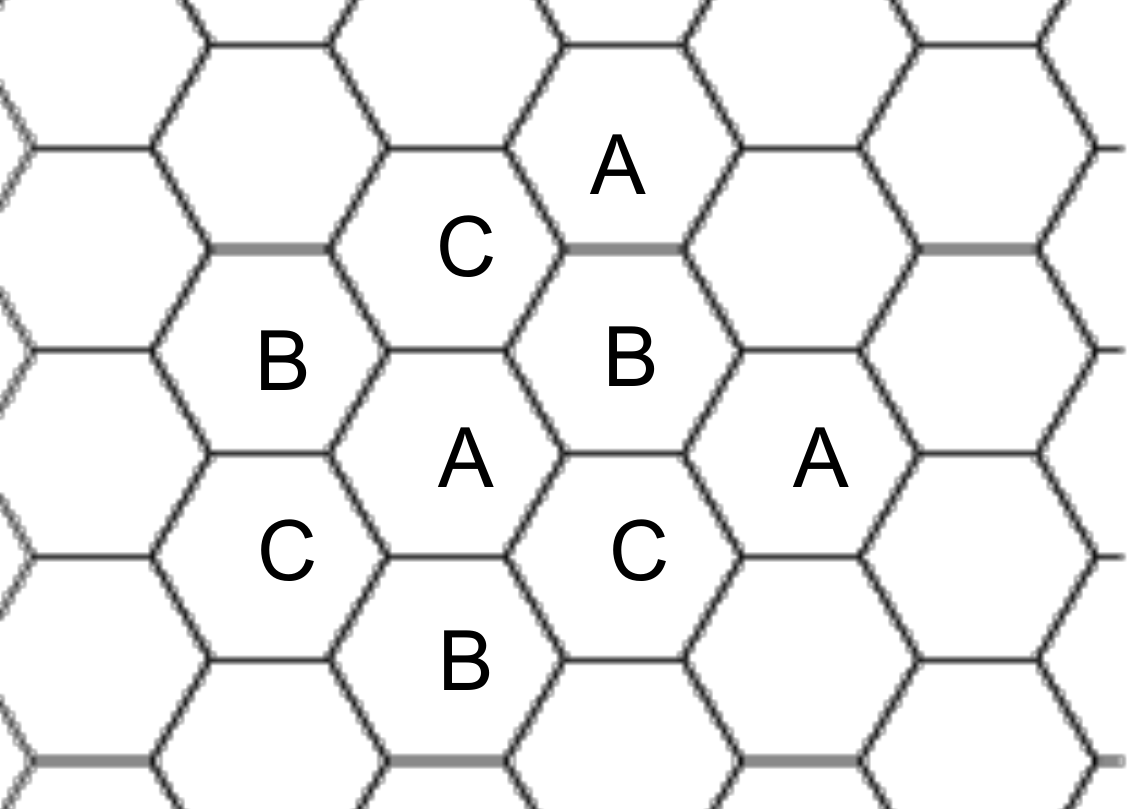}
\caption{Hexagonal lattice showing the three hexagonal sublattices, where each A cell is surrounded by 3 B and 3 C cells, each B by 3 A and 3 C, and each C by 3 A and 3 B.}
\label{lattice}
\end{figure}

The Notch-Delta interaction is an example of juxtacrine (i.e., contact-dependent) signaling. As sketched in Fig. \ref{juxtacrine},  Ligands such as Delta bind Notch receptors and, when this occurs between neighboring cells, leads to the cleavage of the receptor and release of its intracellular domain (NICD). NICD translocates to the nucleus where it transcriptionally up-regulates Notch and down-regulates Delta. The ligand-receptor interaction between molecules on the same cell leads to mutual annihilation with no NICD release~\cite{elowitz}. The combination of  cis-annihilation and NICD-mediated trans-repression is responsible for the observed lateral inhibition of Delta~\cite{collier}. We will use a baseline model~\cite{model} of this process involving three concentrations, $N$ (receptor), $D$ (ligand) and $I$ (NICD),
\begin{eqnarray}
\dot{N}_x & = &  \lambda_N H_+ (I_x)  -N_x \left (k_c D_x + k_t D^{\mbox {\scriptsize ext}} _x  \right) - \gamma N_x \nonumber \\
\dot{D}_x & = &  \lambda _D  H_- (I_x)  -D_x \left (k_c N_x + k_t N^{\mbox {\scriptsize ext}} _x  \right) - \gamma D_x\nonumber \\
\dot{I}_x & = &  k_t \lambda_I N_x D^{\mbox{\scriptsize ext}}_x  -\gamma _I I_x
\label{NDI}
\end{eqnarray}
Here, positions $x$ refer to locations on an hexagonal lattice (see Fig. \ref{lattice}) and the superscript ``ext" refers the average over the six nearest neighbor sites of $x$. The production terms $H_\pm$ corresponding to the aforementioned transcriptional regulation are taken to be Hill functions,
\begin{equation}
H_+(I) 
= 1 + \frac{k_H I
^{\rule[-.3\baselineskip]{0pt}{0pt}{n_+}}}{I_{\strut}^{n_+} + s_0^{n_+}} ; \qquad
H_-(I) 
=  \frac{s_0^{n_-}}{I_{\strut}^{n_-} + s_0^{n_-}} 
\end{equation}
such  that $H_\pm(0)=1$ and $H_+$ is an increasing function that saturates at $1+k_H$, while $H_-$ is a decreasing function that decays to 0 with increasing $I$.
We define a typical set of parameters taken from the literature~\cite{Jagged-Delta}: 
$\gamma =0.1$. $\gamma_I = 0.5$, $n_+ = n_-= 2$, $k_c = 0.1$, $k_t = 0.04$, $\lambda_I =1$, $k_H = 1$, $s_0 = 1$, and primarily focus on the role of $\lambda_N$ and $\lambda_D$.  Nevertheless, we will also have occasion to investigate the effects of changing $k_H$ and $k_t$.

\section{Uniform patterns and the Region of Instability}
We start by considering the simplest type of solution for this system of equations, namely one which is spatially uniform. In general, solving for such a solution requires the simultaneous solution of three nonlinear equations for $N_0$, $D_0$ and $I_0$.
In Appendix A, we show that one can reduce the problem to the solution of one rather complicated equation for $I_0$; this is given as Eq. A5. For general values of the parameters, this last equation must be solved numerically. 

We can gain some analytic insight into the solution space by considering various limiting cases (for full details see Appendices \ref{sec:LargeLambdaD} and \ref{sec:LargeLambdaN}). One convenient such case is that of large $\lambda _D$. In that limit, we expect $D_0$ to be large, which then forces $N_0$ to be small to satisfy the condition arising from the last of Eqns. 1, 
$$
k_t \lambda_I N_0 D_0  = \gamma _I I_0
$$
From the Notch equation, we have in  this limit
$$
N_0 = \frac{ \lambda _N H_+ (I_0)}{(k_c+k_t) D_0}
$$
which then immediately leads to the condition
 \begin{equation}
 I_0 \ = \ {\cal{R}}(I_0) \equiv  \frac{ k_t \lambda_N(1 + k_H \frac{I_0^2}{1+I_0^2}) }{\gamma_I k_0} . \label{eq0}
 \end{equation} with $k_0 \equiv k_c +k_t$ and where we have specialized to the case of Hill functions with $n_\pm= 2$. From  this equation we can determine when there is a unique solution versus when multiple solutions exist. The details of this calculation are presented in Appendix \ref{sec:multiple}. For our purposes here, we note that the range of parameters for which there is a unique uniform solution ranges over all reasonable values of the system parameters.

\begin{figure*}
\includegraphics[width=0.45\textwidth]{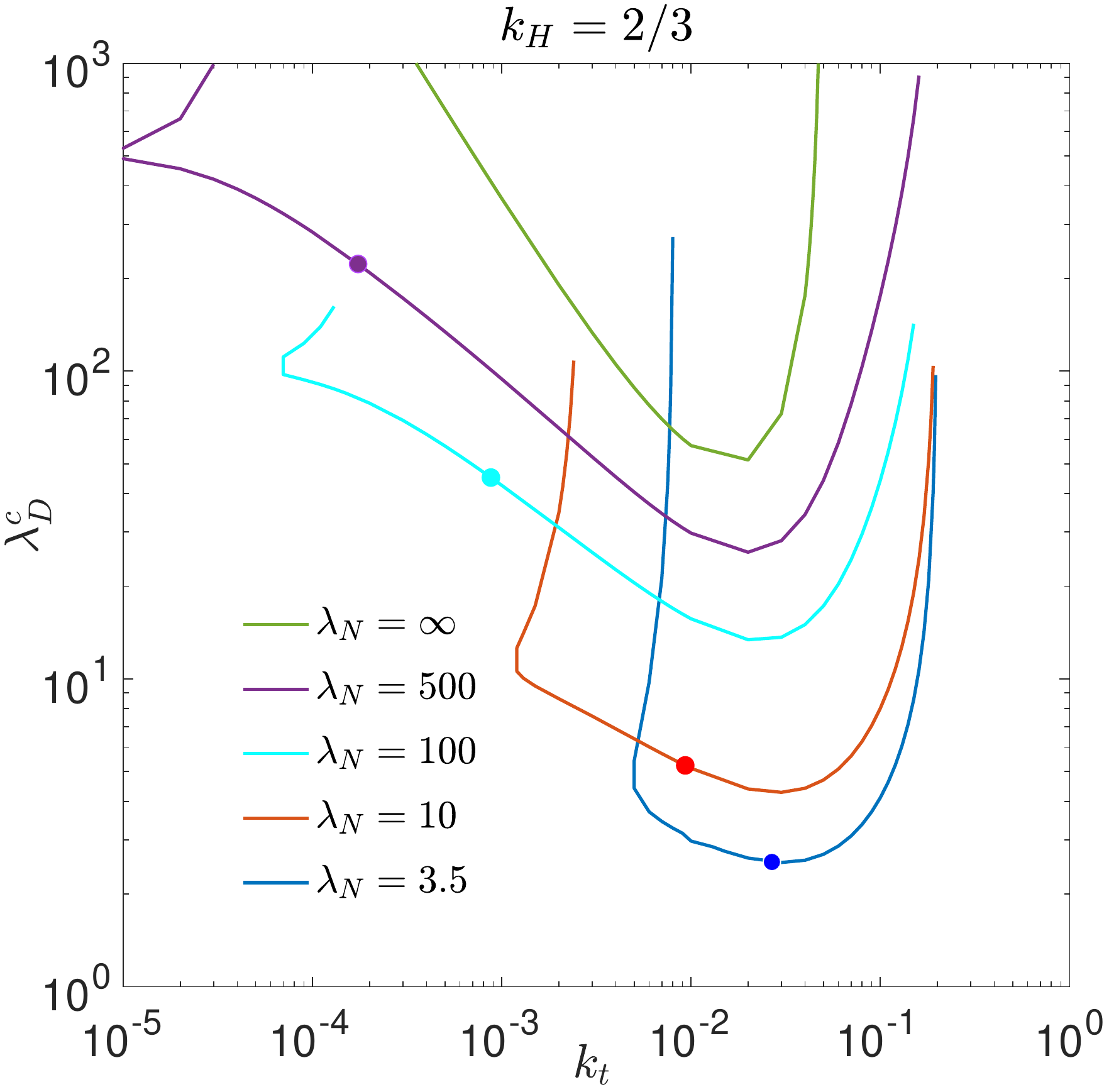}\quad
\includegraphics[width=0.45\textwidth]{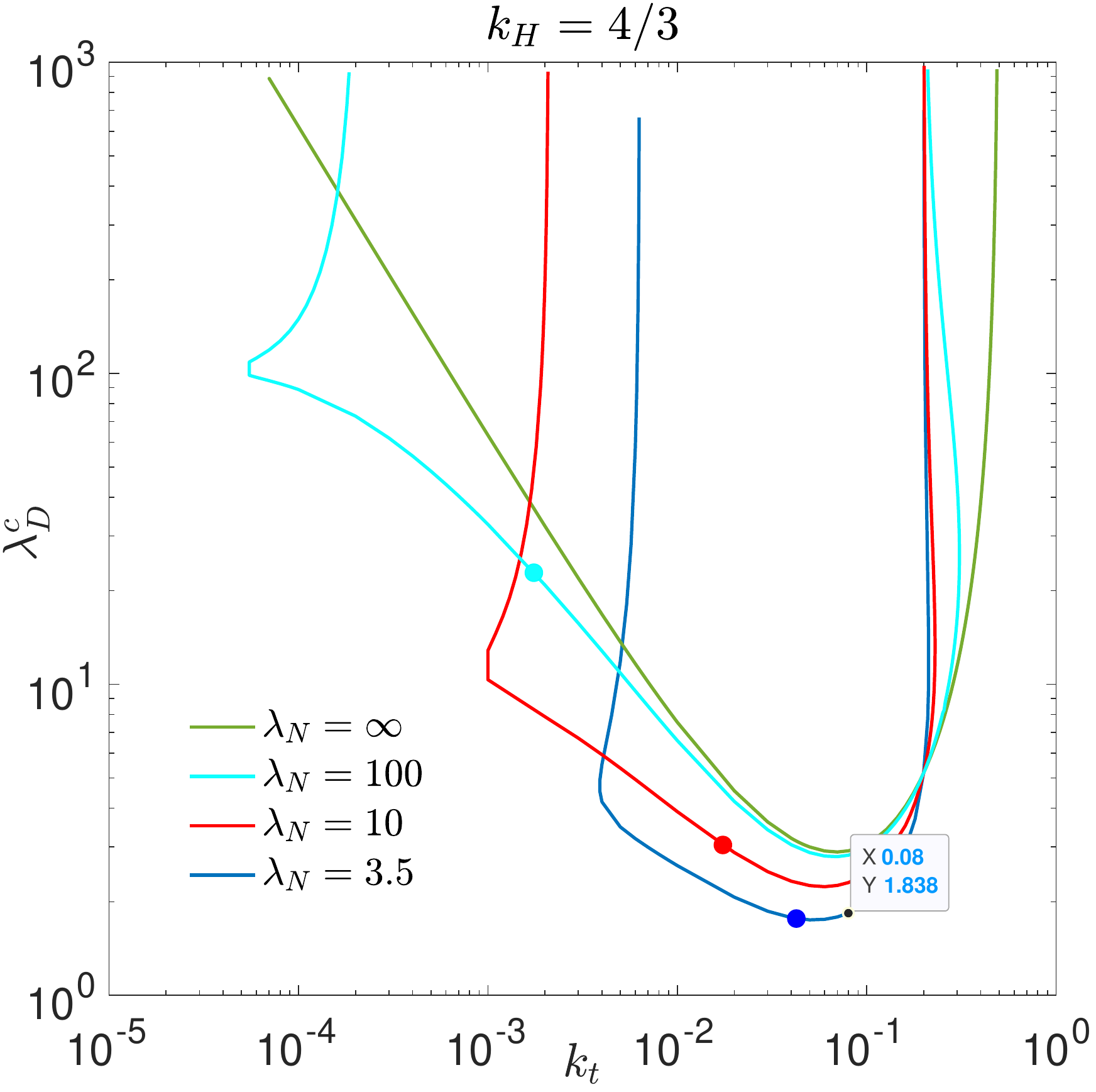}
\caption{The onset of the uniform state instability, giving the critical value of $\lambda_D$ varying $k_t$, for a sequence of (large) values of $\lambda_N$; the plots of the left, right correspond to $k_H=2/3$, $4/3$. All other parameters are as stated in the text: $\gamma =0.1$, $\gamma_I = 0.5$, $n_+ = n_-= 2$, $k_c = 0.1$, $\lambda_I =1$,  $s_0 = 1$.}
\label{figstability}
\end{figure*}

We are interested here in hexagonal patterns on our hexagonal lattice. These patterns arise as the uniform state becomes unstable with respect to spatially varying perturbations. In fact, our numerical experiments on Eq. \eqref{NDI} indicate that the first instability of the uniform state for the set of parameters above is almost always to a hexagonal mode, as that arrangement maximizes the average number of  "satisfied" nearest neighbor inhibitory interactions. This is of course a consequence of studying the model on a  hexagonal grid, as motivated by the biological application. Then, the immediate question is where in parameter space the uniform state is unstable to  a hexagonal pattern.  The condition for this is presented in Eq. \eqref{eq:stab} in Appendix \ref{sec:Reduction}. The region of instability in the $k_t$, $\lambda_D$ plane is presented for various values of $\lambda_N$, both for $k_H=2/3$ in the left panel of Fig. \ref{figstability} and for $k_H=4/3$ in the right panel. 
For each value of $\lambda_N$ considered, the unstable region is that above the corresponding curve, i.e. $\lambda_D>\lambda_D^c(k_t)$.  For each value of $\lambda_N$, the value of $\lambda_D^c$ diverges at two values of $k_c$, with the instability only possible between these two values.  For $k_H=2/3$, the curve $\lambda_D^c(k_t)$
doubles back on itself for low $k_t$, but not near the upper limiting value of $k_c$, whereas for $k_H=4/3$, this doubling back happens on both sides.  This qualitative change in behavior is explained in Appendix \ref{sec:LargeLambdaN}, and occurs at $k_H=1$. Clearly wherever there is a hexagonal instability, a hexagonal solution exists.  However, as we shall see, there is also the possibility of coexistence of a stable uniform solution with a hexagonal solution, so that hexagonal solutions extend outside the region of linear instability of the uniform state.

\section{Hexagonally Ordered Patterns}

We now turn to the construction of hexagonally ordered patterns whose existence (but not stability) is guaranteed by the linear instability discussed above. These patterns are invariant under translation with vectors $\pm 6 \hat{x}$, $\pm 3\hat{x} \pm 3\sqrt{3}\hat{y}$ where $\hat{x}$ and $\hat{y}$ are unit vectors along the coordinate axes and the unit of length is 1/2 the length of one of the hexagonal sides. From Fig. \ref{lattice}, it is clear that the fields everywhere are completely determined by their values on three sub-lattices that we have labeled A, B and C. This means that the entire problem of hexagonally ordered pattern structure (and their stability with respect to modes invariant under translation of the A-B-C unit cell) is reduced to nine coupled ODE's.

This exact mapping is very different than what occurs for more traditional pattern formation problems~\cite{cross} such as for example for convection rolls~\cite{convection}, where the reduction to a set of ODE's is valid only as an approximation near the bifurcation point.  This ODE reduction has been performed previously by Negrete and Oates~\cite{Negrete}  for their simple one-field model of the Notch-Delta system (see below for details on this model), and here it is essential for our analysis of the full realistic model, especially away from the region of the bifurcation. As already shown,   at fixed $\lambda _N$, the uniform solution with the fields taking on the same values on all three sublattices becomes unstable for $\lambda _D>\lambda_D^U(\lambda_N)$  via a transcritical bifurcation, i.e. an intersection with a nonuniform solution.  On this new branch, the respective values of the fields on  two  sublattices (say B and C) are identical, differing from the values on the  remaining (in this case, A) sublattice. This hexagonally structured solution has  a 6-fold hexagonal  symmetry about any site on the different (here, A) sublattice.  The bifurcation is transcritical in general, because of the lack of any symmetry between positive and negative deviations of the fields from their uniform values. As already mentioned and unlike convection, the transition to rolls does not take place at the same parameter value as that to hexagons as the roll pattern necessarily has a different wavelength on the lattice; hence rolls do not compete with hexagons, at least near the bifurcation. For completeness, the 6 coupled equations that govern these ordered hexagonal patterns are given explictly in Appendix \ref{sec:hex}

\section{The Pitchfork Bifurcation and its Unfolding: The Negrete-Oates model}
To understand the nature of the origin of hexagonal patterns in a hexagonal lattice system, Negrete and Oates~\cite{Negrete} introduced a simple one-field model containing the same type of instability. The model
is given by 
\begin{equation}
\dot{u}_n = -u_n - \gamma u_n^3 + \Omega_0 + \epsilon u_\textit{ext}
\end{equation}
where again $u_\textit{ext}$ is the average of $u$ on the 6 nearest-neighbor sites.  For sufficiently negative $\epsilon$, this model has a lattice analog of a negative diffusion constant and hence become unstable to hexagonal patterns. On the three sites of the unit cell introduced above, this system reduces to
\begin{align}
\dot{A} &= -A - \gamma A^3 + \Omega_0 + \epsilon (B+C)/2\nonumber\\
\dot{B} &= -B - \gamma B^3 + \Omega_0 + \epsilon (A+C)/2\nonumber\\
\dot{C} &= -C - \gamma C^3 + \Omega_0 + \epsilon (A+B)/2
\end{align}
They noted the presence of a pitchfork bifurcation at a specific value of the parameters, namely $\epsilon_\textrm{PF} = -2$, $\Omega_\textrm{PF} = 0$. The latter is directly dictated by the symmetry $u\leftrightarrow-u$  of the equations, which of course only is present at $\Omega _0 =0$. It is useful to go through the exercise of working out the bifurcation analysis for this model, as the algebraic structure of our three-field more biologically realistic model is quite similar in structure but more complicated in detail. 

The homogeneous stationary state $A=B=C\equiv H$ satisfies 
\begin{equation}
0 = -H - \gamma H^3 +\Omega_0 + \epsilon H
\end{equation}
with solution $H \approx \Omega_0/(1-\epsilon)$ for small $\Omega_0$.  At the critical value  $\epsilon_c = -2 - 6\gamma H^2$, the stability matrix around the homogenous solution has two zero modes, $(\delta A, \delta B, \delta C)  = (-2,1,1)$ and $(0,1,-1)$ and the nonzero mode $(1,1,1)$, with eigenvalue $-3$. We will be carrying out a weakly-nonlinear analysis in the neighborhood of the co-dimension  two pitchfork bifurcation.point. To obtain this, we write
\begin{align}
A &= H - 2C_1 + \delta_2\nonumber\\
B &= H + (C_1 + C_2) + \delta_2 \nonumber\\
C &= H + (C_1 - C_2) + \delta_2
\end{align}
where $C_1$ and $C_2$ are the ${\cal O}(\Omega_0)$ amplitudes of the zero modes, and $\delta_2$ is an  ${\cal O}(\Omega_0)^3$ correction.
Writing $\epsilon = \epsilon_c + \Delta \epsilon$, where $\Delta \epsilon$ is ${\cal O}(\Omega_0)^2$, and expanding to third order in $\Omega_0$, we obtain three equations for the time-dependent amplitudes $C_1$ and $C_2$ (which vary on the slow time scale $\Omega_0^2$) and $\delta_2$. Eliminating $\delta _2$, we find the two amplitude equations
\begin{align}
\dot{C}_1&=\frac{-C_1 \Delta\epsilon}{2} - 3\gamma C_1^3 - \gamma C_1 C_2^2 + \gamma H (3C_1^2 - C_2^2) \nonumber\\
\dot{C}_2&=\frac{-C_2 \Delta\epsilon}{2} - \gamma C_2^3 - 3\gamma C_1^2 C_2 - 6 \gamma H C_1C_2 
\label{bifur} \end{align}
where the time derivatives refer to variation on the aforementioned slow time scale. It is easy to see that all the terms in our equations are ${\cal O}(\Omega_0)^3$.

We start by looking for time-independent solutions of these equations. One stationary state of the system is the original homogeneous state with $C_1=C_2=0$. There is a pair of solutions with $C_2=0$, with $C_1$ satisfying the quadratic equation
\begin{equation}
0 = - \frac{\Delta \epsilon}{2} + 3\gamma H C_1 - 3\gamma C_1^2 
\label{eqSN}
\end{equation}
 These two solutions emerge  from a saddle-node bifurcation occurring at $C_1= H/2$, $\Delta\epsilon=3\gamma H^2/2$. This lies on the stable side of the transition and thus  the saddle node bifurcation precedes the instability of the homogenous state as $\epsilon$ is decreased..  As $\Delta\epsilon$ further decreases from its saddle-node value, one of the solution branches has increasing $C_1$, while the other approaches $C_1=0$, i.e., the homogeneous solution, intersecting it  at $\Delta\epsilon=0$, the location of the homogeneous instability. It then crosses over to $C_1<0$, so that here it has  ``polarity'' opposite to that of the other branch emerging from the saddle-node bifurcation that had $C_1>0$ and increasing.

  There are other stationary solutions, having $C_2 \ne 0$.  Solving the $\dot{C}_2=0$ equation for $C_2$ and substituting in the $\dot{C}_1=0$ equation yields 
 $C_1=-x/2$, where $x$ is the value obtained above for $C_1$ when $C_2=0$. Substituting this back into the equation for $C_2$ then reveals $C_2 = \pm 3x/2$.
 For these solutions, the leading order value of $A - H$ is $-2C_1 = x$, which was the leading order value of $B-H=C-H$ in our original solution. In addition, either $B-H$ or $C-H$ (depending on the sign of $C_2$) equals $x$ as well, with the other equalling $-2x$, which was the value of $A-H$ in the original solution.  Thus, all these new  solutions are simply the previous solutions translated to be centered on $B$ or $C$, instead of $A$.  At the linear level, these new solutions are the linearly combinations $-1/2[(-2,1,1) \pm 3(0,1,-1)]$ of the two zero modes of the homogeneous solution.  This crossing of this inhomogeneous solutions with the original homogeneous solution represents a two dimensional  transcritical bifurcation of the homogeneous solution; two-dimensional here because of the two zero modes of the homogeneous solution.  Precisely at $\Omega_\textrm{PF}$, i.e., $\Omega_0=0$, the saddle node merges with the transcritical point and the two branches meet symmetrically at $\Delta\epsilon=0$, i.e., $\epsilon_\textrm{PF}$, as indicated by the vanishing of the term linear in $C_1$ in Eq. \eqref{eqSN}.

 We can also easily calculate the stability spectrum of the non-trivial patterned states of this reduced system. The homogeneous solution has a degenerate pair of  modes, with growth rates $\Omega_1=\Omega_2=-\Delta\epsilon/2$, so that is stable below the transition, $\Delta\epsilon>0$ and unstable above. Focusing on the $C_2=0$ inhomogeneous solution, it has two eigenvalues, $\Omega_1 = 3\gamma C_1 (H- 2C_1)$ and $\Omega_2=-9\gamma C_1 H $.  From this we can see that (for $\Omega_0>0$) on the positive branch both modes are stable. $\Omega_1$ crosses 0 at the saddle node, and on the second branch we have 1 stable and one unstable mode.  When the second of the non-uniform branches crosses the homogeneous solution at the transcritical point, both $\Omega$'s cross zero and we remain with one stable and one unstable mode, switched compared to those on the previous side. The story for the other, shifted, solutions is of course the same.
 This picture is more intricate than for the standard co-dimension one pitchfork, where the stable solution on one side of the bifurcation gives rise to one unstable and two symmetry-related stable solutions on the other side.
  
 \section{The Pitchfork Bifurcation and its Unfolding: The Biological Notch-Delta model}

The shared symmetry structure with the Negrete-Oates model above suggests and numerical exploration of Eq. \eqref{NDI} confirms that our more realistic Notch-Delta system also possesses a co-dimension 2 pitchfork bifurcation. For our standard parameters the pitchfork occurs at  at $\lambda_N^\textit{PF}=2.7582488$, $\lambda_D^\textit{PF}=1.93112$.  Specifically, for $\lambda_N^\textit{PF}$, as is the case for general $\lambda_N$,  only the uniform solution exists for $\lambda_D<\lambda_D^U$ and it is stable.  At $\lambda_D^U=1.93112$, two additional solutions are born, one a ``hexagon" (by definition, a solution where high $D$ is surrounded by high $N$) and one an ``anti-hexagon" (high $N$ surrounded by high $D$).  What is interesting is that this point occurs not only for physically possible (i.e. positive) values of the parameters, but within the range of parameter values determined (at least roughly) by experiment \cite{elowitz}.  Thus the unfolding of the bifurcation gives us detailed information about the pattern formation possibilities realizable in real physiological settings.

\begin{figure}
\includegraphics[width=0.45\textwidth]{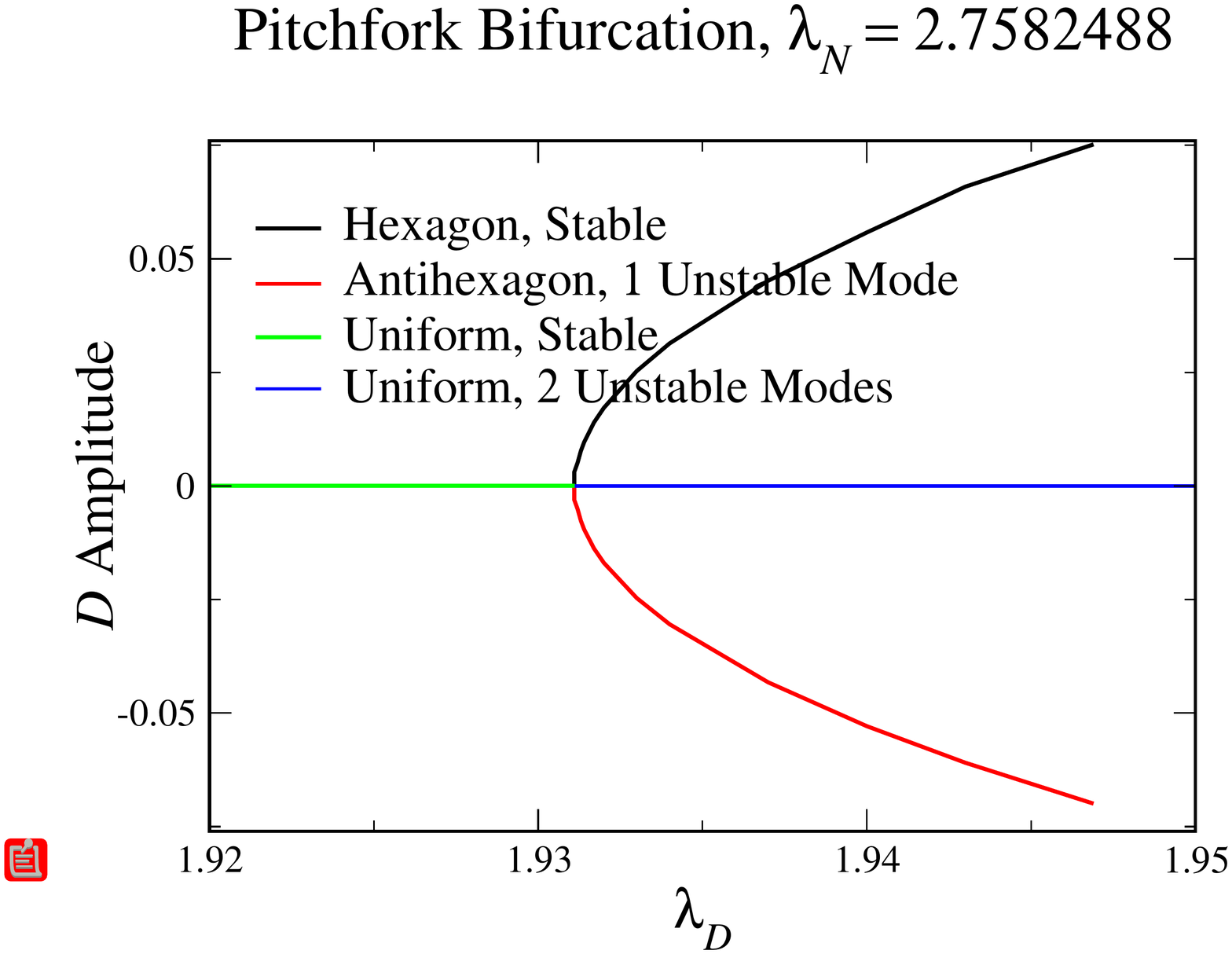}
\includegraphics[width=0.45\textwidth]{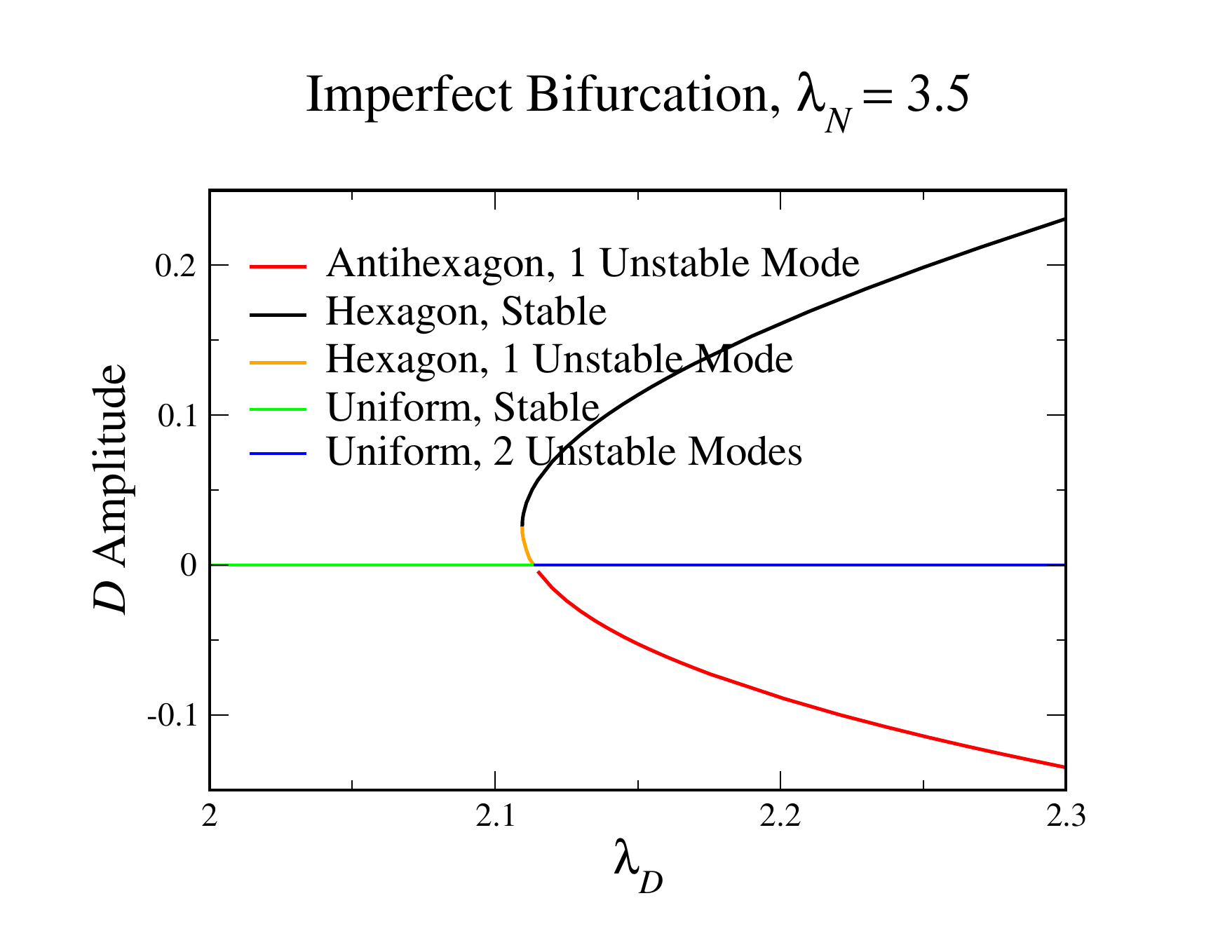}
\caption{Bifurcation diagrams: A)  for the critical $\lambda_N^\textit{PF}=2.7582488$ for which there is a pitchfork bifurcation; B) For $3.5=\lambda_N>\lambda_N^\textit{PF}$ where there is an imperfect bifurcation. } \label{bifurc}
\end{figure}

Many of the features in the vicinity of the pitchfork bifurcation of the Negrete-Oates model carry over into our more complicated system.  Specifically, it can directly be shown by numerical analysis of the 9-dimensional reduced system that the uniform state has 2 (degenerate) unstable modes above the critical $\lambda_D$.  The emerging hexagon branch is stable, whereas the anti-hexagon has one unstable mode.  The instability is with respect to a mixed mode (defined as a mode with all three sublattices having different values) which converts the anti-hexagon to a shifted hexagon.  This overall structure is shown in Fig. \ref{bifurc} A, and indeed recapitulates the structure determined for the Negrete-Oates model via the amplitude equation analysis.
\begin{figure}
\begin{center}
\includegraphics[width=0.5\textwidth]{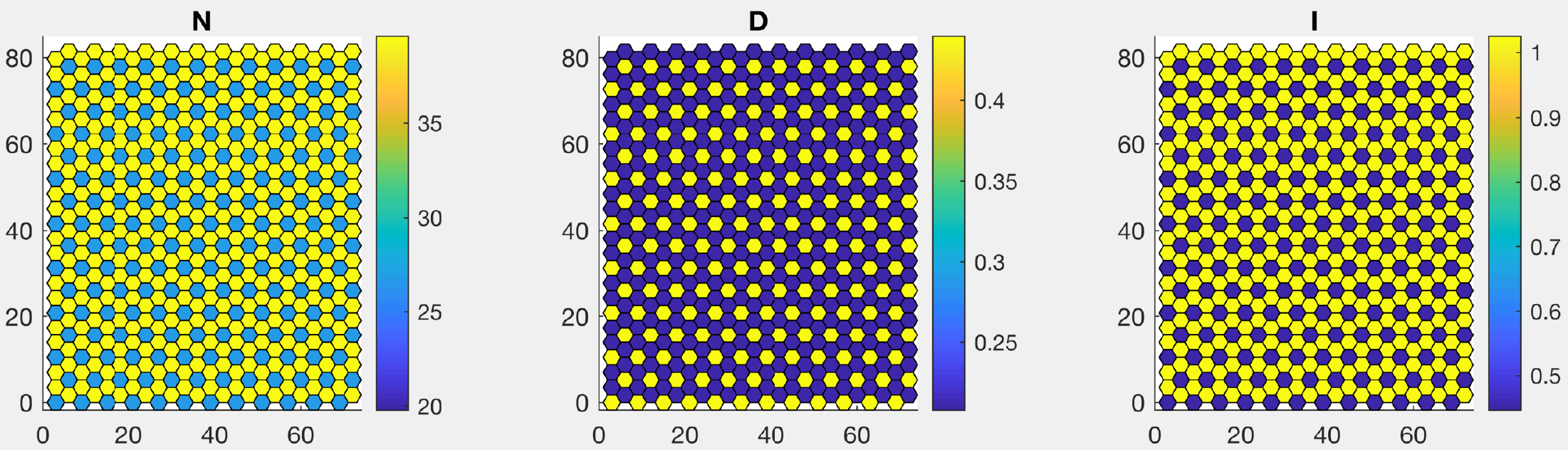}
\end{center}
\caption{A stable hexagon solution for $\lambda_N=3.5$, $\lambda_D=2.3$.}
\label{hexfig}
\end{figure}

At all other values of $\lambda_N$, the pitchfork breaks up into a transcritical bifurcation and a saddle-node (see Fig. \ref{bifurc}B), also as in the Negrete-Oates model.  Again by direct numerical solution, we find that for $\lambda_N>\lambda_N^\textit{PF}$, the uniform solution undergoes a transcritical bifurcation with a unstable anti-hexagon (with respect to a mixed-mode perturbation) on the high $\lambda_D$ side and an unstable (with respect to a pure-mode) hexagon on the low $\lambda_D$ side.  The unstable hexagon then undergoes a saddle-node bifurcation, rendering the hexagon stable; this stable branch then continues on as $\lambda_D$ increases.  For $\lambda_D$ smaller than the saddle-node value, no patterned solution exists. Hence, there exists a range of parameters for which a stable hexagon coexists with the stable uniform solution, a range which widens as $\lambda_N$ increases; we will return to this point below.  For example, for $\lambda_N=3.5$, the transcritical bifurcation in which the uniform state goes unstable is at $\lambda_D=2.11356$, whereas the saddle node bifurcation is at $\lambda_D=2.1097$. An example of a stable hexagon solution for $\lambda_N=3.5$, $\lambda_D=2.3$ is shown in Fig. \ref{hexfig}.
\begin{figure}
\includegraphics[width=0.45\textwidth]{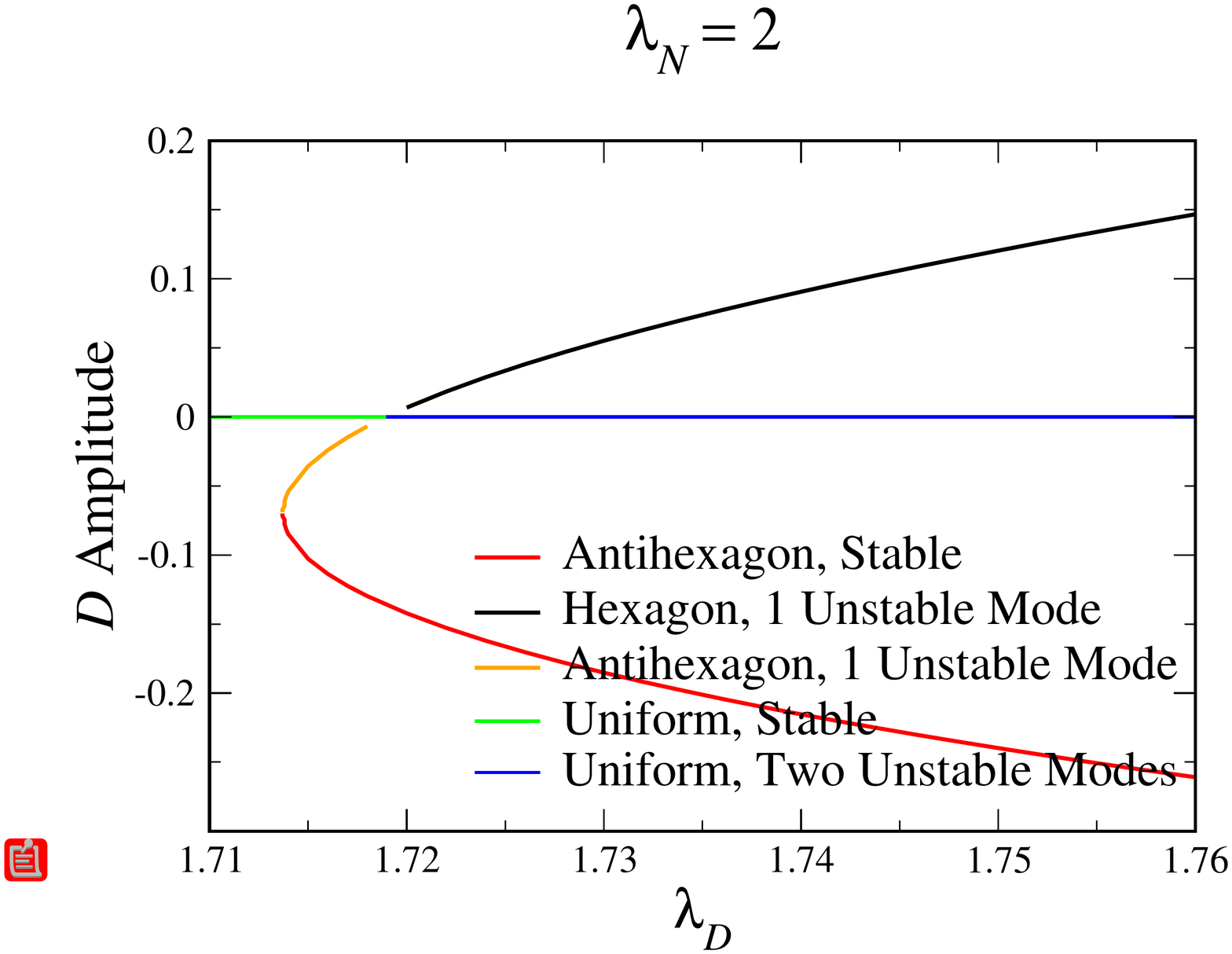}
\includegraphics[width=0.35\textwidth]{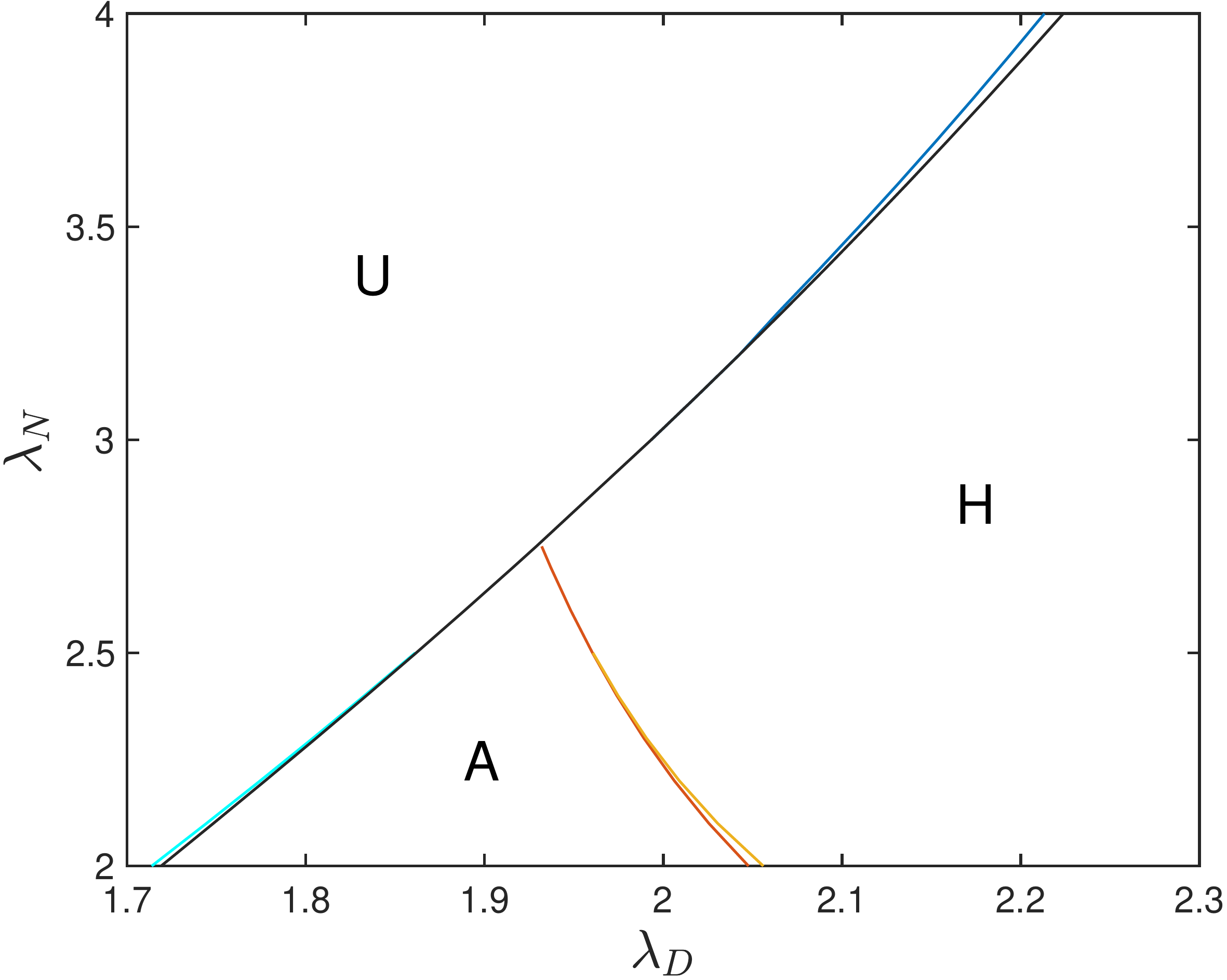}
\caption{A. Bifurcation diagram for  $2.0=\lambda_N<\lambda_N^\textit{PF}$ . B. Phase diagram in the $\lambda_D$, $\lambda_N$ plane indicating stable phases: U (uniform), H (hexagon), A(anti-hexagon) regions; the red line corresponds to the mixed-mode instability. Note that there are very tiny, almost invisible (on this scale), regions of two-phase coexistence that exist between lines delineating boundaries of the strictly monophase regions; see, for example, the separation between two lines, corresponding  to the loss of stability of the U (rightmost, black) and H (leftmost, blue) states, at the top right.  All other parameters are standard.} \label{transcritical}
\end{figure}

A similar bifurcation structure appears for $\lambda_N<\lambda_N^\textit{PF}$, where now the stable hexagon lies to the right of the transcritical point and the unstable anti-hexagon lies to the left, and it is the one that undergoes a saddle-node bifurcation (see Fig. \ref{transcritical}A).  The anti-hexagon is born with 1 unstable mode at the transcritical point and turns stable at the saddle-node bifurcation. However, unlike what happened in the Negrete-Oates model, cite{Negrete}. the stable anti-hexagon branch loses stability to a mixed-mode perturbation; this instability leads to a new pitchfork bifurcation, which is a result of the $B/C$ symmetry breaking. The hexagon, on the other hand, is born with one unstable mode and subsequently becomes stable, also as a result of a mixed-mode pitchfork bifurcation. An overall diagram of the stable phases as a function of the two parameters $\lambda _N$ and $\lambda _D$ is presented in Fig. \ref{transcritical}B  The mixed-mode solution branch arising from the hexagon bifurcation is the same solution which arises from the anti-hexagon bifurcation.  For example, at $\lambda_N=2$, the hexagon becomes stable at $\lambda_D=2.0477$ and  the anti-hexagon becomes unstable at $\lambda_D=2.056$. Thus, there is a very small coexistence region between the hexagon and anti-hexagon solutions. Again, the only solution that survives stably to higher values of $\lambda _D$ is the hexagon. This is in accord with the general biological rule given above that the high Delta cells are surrounded by high Notch cells sufficiently far from $\lambda_N^\textit{PF}$ and its associated $\lambda_D^U$. From the physics perspective, the explicit lack of symmetry between Notch and Delta as reflected in this model is not eliminated by working close to the co-dimension two bifurcation since the accidental symmetry at this point affects only the leading order term in the amplitude equation, not any of the higher-order ones. This feature is not captured by the simpler one-field model where the model has an exact symmetry at $\Omega _0 =0$.

We can again use weakly non-linear bifurcation theory to flesh out these numerical findings, working in the immediate vicinity of the pitchfork bifurcation, 
The bifurcation analysis for this more complicated system of 9 equations can be performed, and after eliminating the seven fast modes (as opposed to just one previously), we get  a set of  amplitude equations the two slow modes, parameterized by $\delta_{B,C} = N_{B,C} - N_0$, where $N_0$ is the homogeneous Notch level. Doing this, we get exactly the same bifurcation equations structure as in Eq. \ref{bifur} with the role of the symmetry breaking parameter $\Omega _0$ played by $\delta\equiv\lambda_N - \lambda_N^\textit{PF}$ and the role of the other bifurcation parameter $\Delta\epsilon$ replaced by $\epsilon\equiv (\lambda_D-\lambda_D^c)$ where  $\lambda_D^c$ is the location of the instability of the homogeneous solution. Near $\lambda_N^\textit{PF}$, 
\begin{equation}
\lambda_D^c \approx 1.931 + 0.262 \delta - 0.0236 \delta^2 +  0.000676 \delta^3
\end{equation}
so that $\lambda_D^c$ increases with $\lambda_N$. We can write the final amplitude equations as  
\begin{align}
\dot{C}_1&=\alpha  C_1 + \sigma (3C_1^2 - C_2^2) + \mu  C_1(3C_1^2 +  C_2^2)  \nonumber\\
\dot{C}_2&=\alpha C_2  - \sigma(6C_1C_2 )+ \mu C_2( C_2^2 + 3 C_1^2)
\label{eq:bifur}
\end{align} 
In term of these bifurcation parameters $\delta$ and $\epsilon$, the amplitude equation coefficients are
\begin{equation}
\alpha = 0.04440\, \epsilon; \; \sigma = 0.000382\, \delta; \; \mu = - 0.0003063 .
\end{equation}
This system is of course identical in structure to that we received in the Negrete-Oates model.  The new information is the connection of the perturbations $C_1$, $C_2$ to the deviations from homogeneity of the physical fields $N$, $D$, and $I$.

The analysis of these equation thus follows directly from our previous analysis. Here, the saddle-node point is at $\epsilon = -0.00807\, \delta^2$, again below the transition, with $C_1^{\textit{SN}} = 0.6244\,\delta$, so that for $\delta>0$ the 
$B$ and $C$ sites has high Notch, so that the $A$ site has low Notch.  From the solution for the fast modes, we have $D_A = D_0 +  0.0276\, C_1$, so the $A$ site has high Delta.  Thus the saddle-node solution is what we entitle a ``hexagon" solution, which of course a distinction that is meaningless in the one-field model.  On the other side of the transcritical point, $C_1$ changes sign, and so the $A$ site there has low Notch and high Delta, i.e. an anti-hexagon.  For $\delta<0$, i.e. $\lambda_N<\lambda_N^\textit{PF}$, things are reversed, and the saddle node has $C_1<0$, so that the saddle-node solution is an anti-hexagon, and the solution on the other side of the transcritical point is a hexagon. This is of course consistent with our detailed numerical findings.

Checking the stability, both homogeneous modes have growth rate $\alpha$, and so are stable for $\lambda_D < \lambda_D^c$ and unstable above the transition. For the $\delta>0$ hexagon, one mode, with growth rate $\Omega_1$ is stable on the upper branch $C_1 > C_{1,\textit{SN}}$ and unstable for $C_1$ below the saddle node. The other mode, with growth rate $\Omega_2$ is stable both above and below the saddle node. Thus, the hexagon on the upper branch is stable and on the lower branch is once unstable.  Across the transcritical point, the two modes switch signs, and the antihexagon also has one unstable mode. For $\delta<0$ on the other hand, the bottom antihexagon is stable and the top antihexagon and the hexagon are once unstable.  Precisely at $\lambda_N=\lambda_N^\textit{PF}$, $\Omega_1$ and $\Omega_2$ both vanish and one has to go to higher order to see that the antihexagon is the unstable solution. Also, the instability of the anti-hexagon to the mixed mode is not present to this order of the amplitude equation analysis.  

\section{Exotic Solutions}

The aforementioned exact mapping of the ordered pattern equations to a coupled ODE system allows as well for the analytic understanding of a surprising type of pattern not heretofore investigated. If
we solve our system at a low value of $k_c = 10^{-4}$, with $\lambda_N=3.5$, $\lambda_D=10$ and all other parameters at standard values, we find a hexagonal structure with $N_A = 52.7 > N_B = 51.0$, while as usual $D_A= 0.558>D_B = 0.215$,. We refer to this type of solution as ``high-high", as cells with high Delta also have high Notch. However, we did not find any evidence for this possibility over the range of cis-inhibition parameters proposed by Sprinzak, et al. for typical developmental processes; this is shown in a numerically computed phase diagram (Fig. \ref{lowkC}). To better understand this diagram, we show in Appendix \ref{HH} how one can analytically derive the boundary between regular hexagons and high-high  solutions in the large $\lambda _D$ limit. Specifically, an accurate approximation for the vertical line in the figure is $k_c ^* = k_t/2(1+k_H)$, which equals  $0.01$ with our typical choices. This result shows that hexagonal patterns are not dependent on having high cis-inhibition but that the anti-correlation between Delta and Notch cannot be taken for granted in its absence.
\begin{figure}[t]
\centering
\includegraphics[width=0.75\columnwidth]{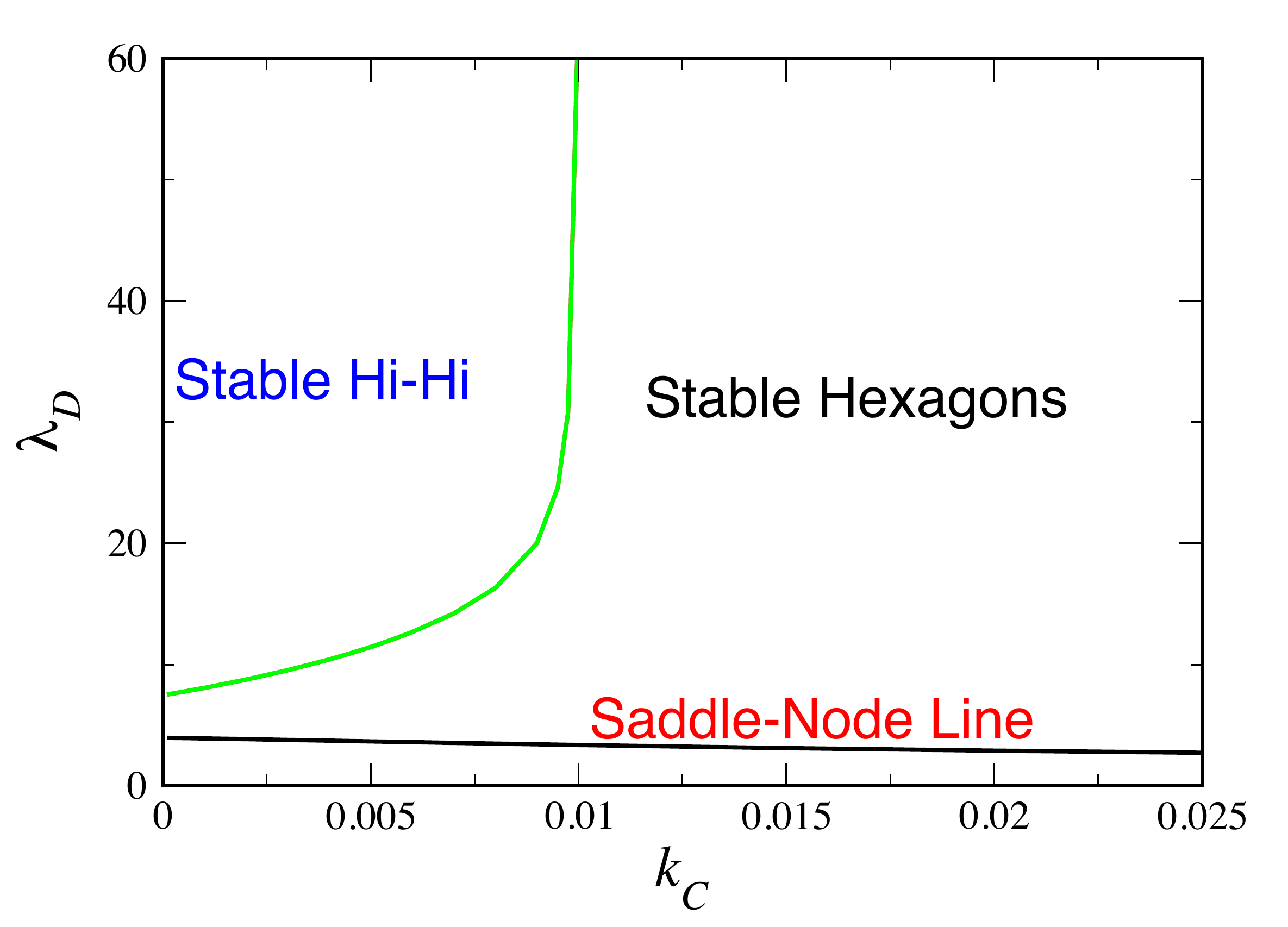}
\caption{Phase diagram in the $k_c$, $\lambda_D$ plane showing the transition from the uniform state (low $\lambda_D$ to hexagonal patterns (Saddle-Node Line) as well as the smooth change from the normal to the high-high pattern. $\lambda_N=3.5$. All other parameters standard.}
\label{lowkC}
\end{figure}

\section{Initial Value Problem} 
The existence of stable ordered hexagons leaves open the question of how these patterns can be generated in the noisy biological system with plausible initial conditions~\cite{barkai, prl}. In particular, it is easy to check numerically that, starting with no pattern for a set of parameters for which the uniform state is linearly unstable,  the presence of noise, either in the initial data or in the time evolution, will lead to disordered states with many domain boundaries between hexagon patterns centered on different sublattices. One way out is based on the fact we have shown above that there could exist a parameter range for which there is a subcritical bifurcation to stable hexagons in which case a local perturbation which nucleates the pattern can spread in an ordered manner; this is a standard scenario in many non-living systems~\cite{cross}. Crucially, the bifurcation analysis suggests that for biological systems studied to date, there is no significant range of physiologically relevant parameters where propagation would occur into a metastable state. Intuitively, we believe that most biological systems exhibit insufficient parameter control and too high a level of stochasticity for this to be a robust strategy. In some, coupling to additional components could alter the bistability range. There could also be more complex biological mechanisms that, for example, would provide downstream checks that prevent neighboring cells from both developing the same phenotype even if there is some initial defect in the Notch-Delta structure~\cite{robust}. 

\begin{figure}
\begin{center}
\includegraphics[width=0.15\textwidth]{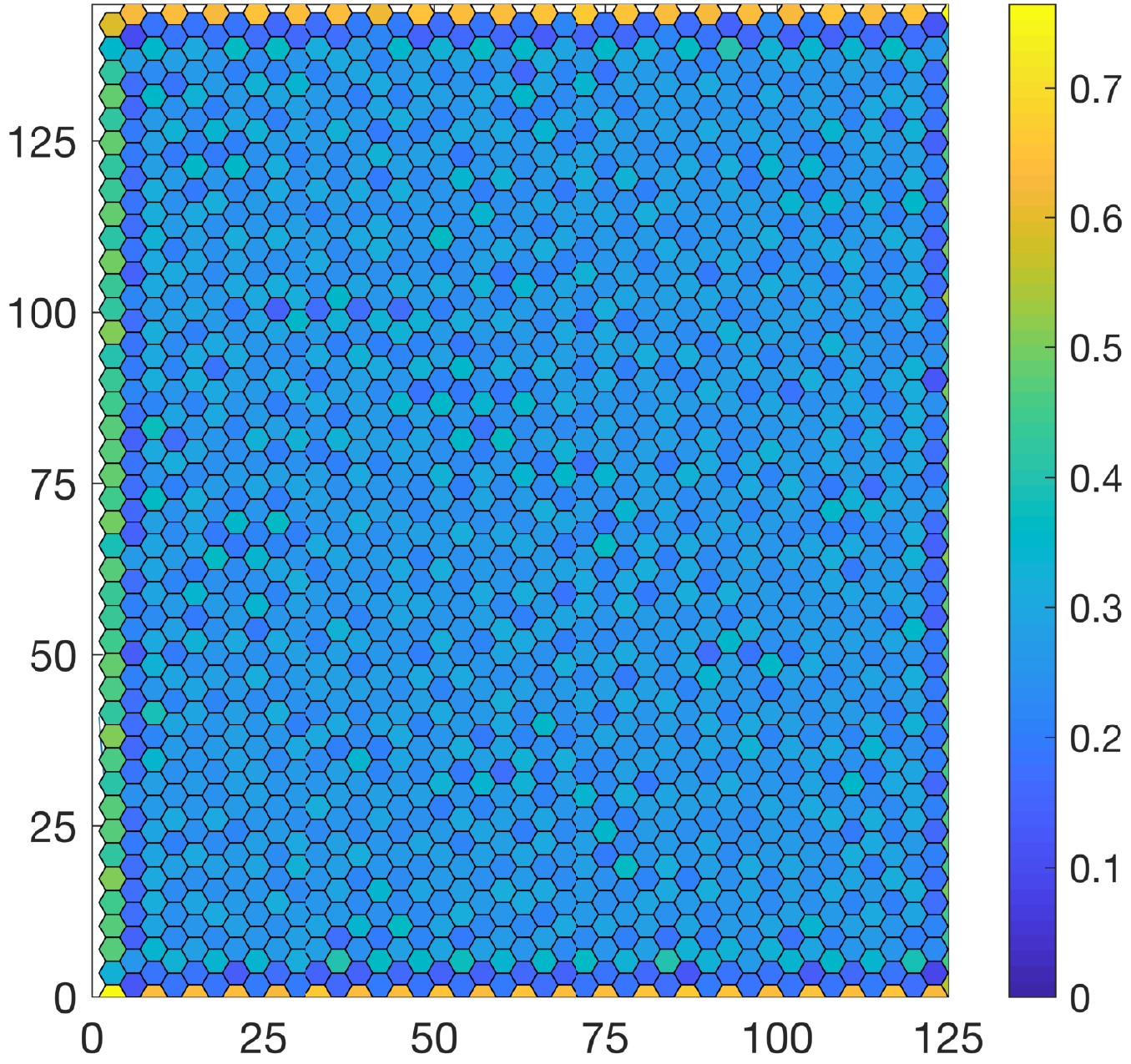}
\includegraphics[width=0.15\textwidth]{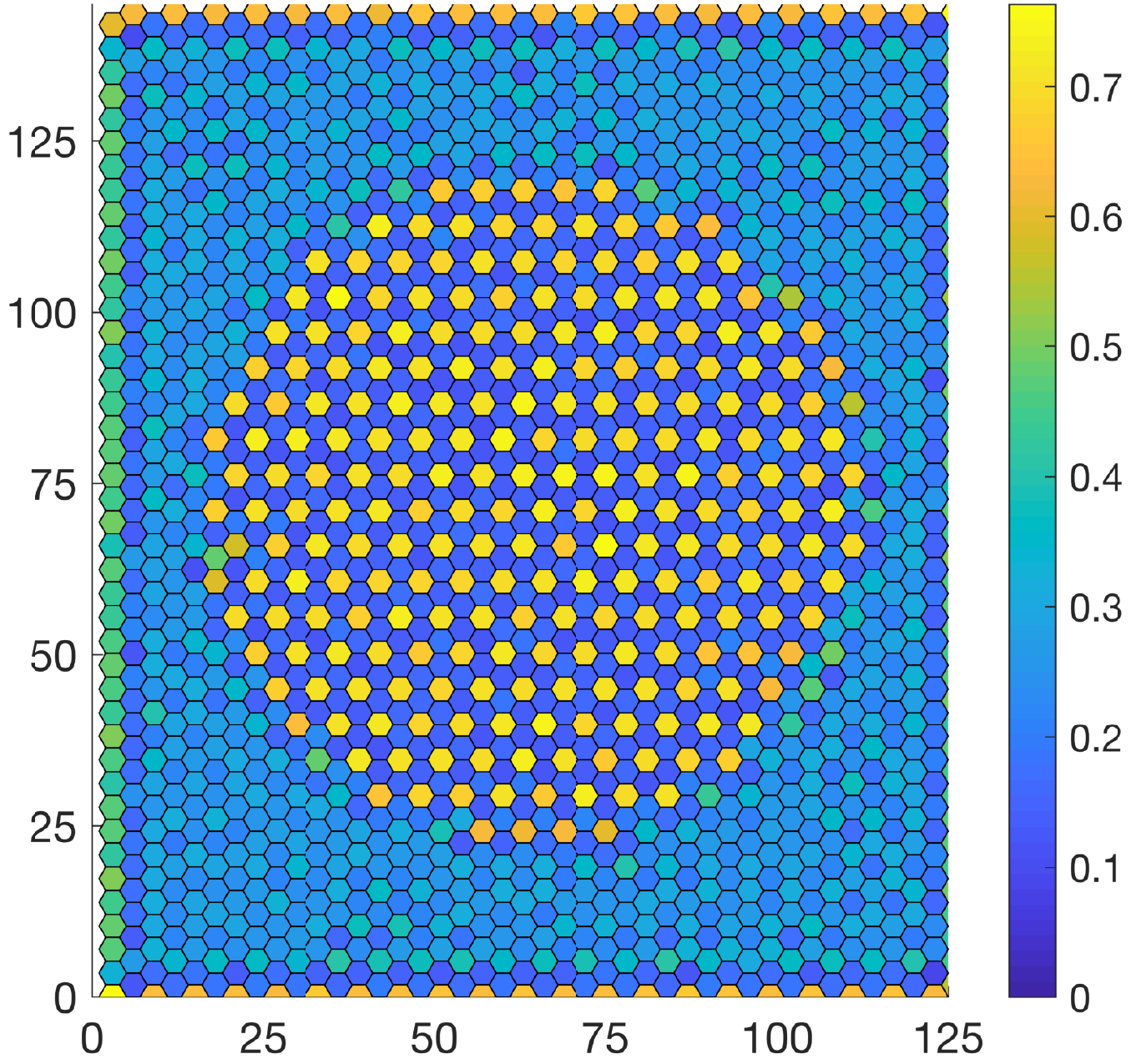}
\includegraphics[width=0.15\textwidth]{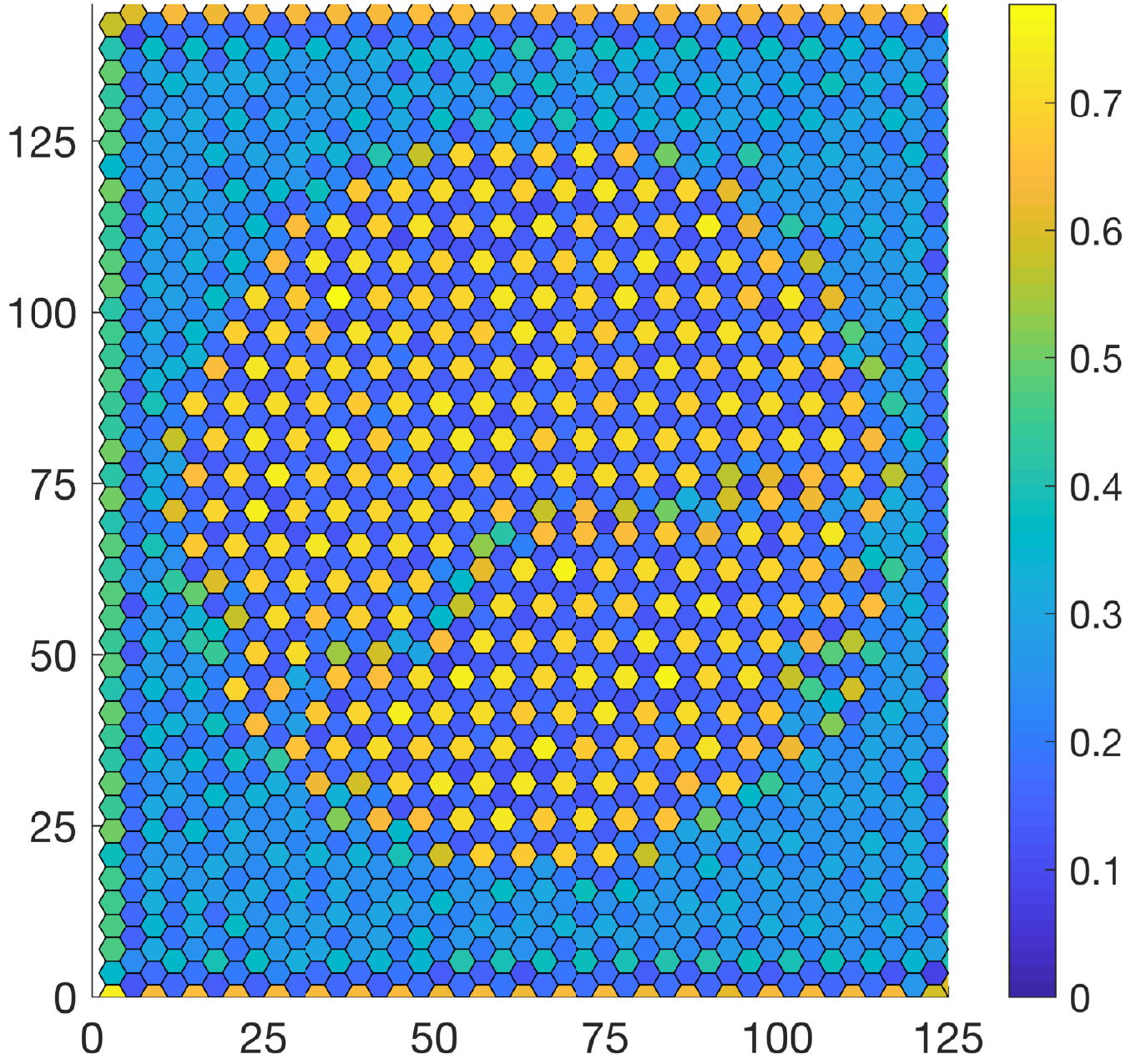}
\end{center}
\caption{The pattern (of $D$, the patterns of $N$ and $I$ are similar) created by a radially propagating initiation wave raising $\lambda_D$ from its initial value of $2$ to a final value of $3.5$. Left panel: The system before the arrival of the wave.   The bulk variation in $D$ is caused by a quenched 3 percent Gaussian variation in $k_t$ from cell to cell. Middle panel: The ordered pattern resulting from a slow ($v=0.015$ cells/s) initiation wave. Right panel: The multi-domain structure resulting from a fast ($v=0.05$ cells/s) initiation wave. $\lambda_N=3.5$. }
\label{fig:Wave}
\end{figure}

A more physics-based possibility is that the system is not put all at once into the unstable state. Rather, we imagine that the system is initially in a regime of parameter space for which the uniform state is stable. Then some external mechanism induces a propagating wave, behind which the  parameters are in the unstable region. To exhibit this possibility, we assume that only $\lambda_D$ is affected by this wave, and  $\lambda_D = 2 $ ahead of the wave and $\lambda_D=3.5$ behind the wave.
We do not concern ourselves here with the origins or dynamics of this initiation wave, and rather choose a standard {\em{tanh}} waveform,  and vary the wave speed $v$. In this regard, our suggestion differs from that of Ref. \cite{pennington2010switch,Shraiman}, who start with a bistable system with two uniform states - in our proposal, the bistable dynamics is not intrinsically related to the Notch-Delta dynamics. In Fig. \ref{fig:Wave}, we show a pair of simulations of our model augmented by quenched noise. The wave in $\lambda _D$ propagates radially outward from an initial point creating an expanding region inside of which the system exhibits Notch-Delta patterning. At large $v$, the parameter shift is essentially instantaneous over a large spatial region  and the noise nucleates incommensurate patterns in  different parts of the lattice, leading to obvious defects. If the parameter wave is slowed down, the leading edge of the pattern has sufficient time to align itself with the preceding rows before having to itself act as a template for the next radial row. One can do this as well in a planar geometry, as is suggested by some biological data~\cite{eyes}. In some sense, all we have done is transfer the problem to one of creating a bistable system responsible for the parameter dependence. But, we assert that this is relatively easy to accomplish and that decoupling the patterning aspect from the bistable aspect (i.e., the Notch system is not bistable at the physiological parameters) is a robust approach to the elimination of defects.

%

\section{Summary}
In conclusion, we have studied the problem of ordered pattern formation for the realistic Notch-Delta system by mapping it  to a 9 degree of freedom ODE system. This mapping facilitates both numerical and analytical progress.  We have shown how the presence of a pitchfork bifurcation value $\lambda_N^\textit{PF}$ close to the physical relevant parameter range organizes the high-Delta centered hexagon pattern as well as the high-Notch centered antihexagon pattern and guarantees that the former is the generic stable structure. Thus, models built on our current understanding of molecular mechanisms do help explain this recurring feature of tissue development. The importance of this is highlighted by the fact that outside the physical range of parameters, alternative correlations between Notch and Delta are possible. Furthermore, we have seen that creating a perfect pattern is a significant challenge in the vast majority of parameter space where the transition from the uniform state to the patterned state is second order.  Lastly, we demonstrated that coupling a parameter to an initiation wave could provide a way to meet this challenge.

\begin{acknowledgments} ET and DAK acknowledge the support of the United States-Israel Binational Science Foundation, Grant no. 2015/619. HL acknowledges the support of the NSF grant no. PHY-1605817.  ET acknowledges useful conversations with David Sprinzak. 
\end{acknowledgments}
\bibstyle{apsrev4-2}
\bibliography{notch}

\onecolumngrid
\appendix
 \renewcommand{\thefigure}{\thesection\arabic{figure}} 
\section{Uniform State Reduction to a Single Equation\label{sec:Reduction}}
\setcounter{figure}{0} 
When discussing uniform steady-states, it is convenient to reduce the set of three equations to a single nonlinear equation. For simplicity, we will specialize the following discussion to the case of the Hill coefficients $n_\pm$ equal to 2. We first introduce the notation
\begin{align}
\tilde{\lambda}_N &= \lambda_N \left(1 + k_H\frac{I_0^2}{1 + I_0^2}\right); \nonumber\\
\tilde{\lambda}_D &= \frac{\lambda_D}{1+I_0^2} .
\end{align}
In terms of these, we have
\begin{align}
N_0 &= \frac{\tilde{\lambda}_N}{\gamma + k_0D_0} ,\nonumber\\
D_0 &= \frac{\tilde{\lambda}_D}{\gamma + k_0 N_0}, \nonumber\\
I_0 &= \frac{k_t \lambda_I}{\gamma_I} N_0 D_0 ,
\end{align}
where we have also introduced
\begin{equation}
k_0 \equiv k_c + k_t .
\end{equation}
Solving for $N_0$ and $D_0$ yields
\begin{align}
N_0 &= \frac{-k_0\left(\tilde{\lambda}_D - \tilde{\lambda}_N\right) - \gamma^2 + \sqrt{\left[k_0\left(\tilde{\lambda}_D + \tilde{\lambda}_N\right) + \gamma^2\right]^2 
- 4k_0^2 \tilde{\lambda}_D\tilde{\lambda}_N}}{2 k_0 \gamma} ,\nonumber \\
D_0 &= \frac{k_0\left(\tilde{\lambda}_D - \tilde{\lambda}_N\right) - \gamma^2 + \sqrt{\left[k_0\left(\tilde{\lambda}_D + \tilde{\lambda}_N\right) + \gamma^2\right]^2 
- 4k_0^2 \tilde{\lambda}_D\tilde{\lambda}_N}}{2 k_0 \gamma} ,
\end{align}
so that $I_0$ satisfies
\begin{equation}
I_0 = \frac{k_t \lambda_I}{\gamma_I} \frac{-k_0\left(\tilde{\lambda}_D + \tilde{\lambda}_N\right) - \gamma^2 - \sqrt{\left[k_0\left(\tilde{\lambda}_D + \tilde{\lambda}_N\right) + \gamma^2\right]^2
- 4k_0^2 \tilde{\lambda}_D\tilde{\lambda}_N}}{2 k_0^2} \equiv {\cal R}(I_0) . \label{eqI0}
\end{equation}
We next consider the stability condition for a hexagonal perturbation, which in line with the Negrete-Oates model has the form $N_A=N_0 + \delta_N, N_{B,C} = N_0  - \frac{1}{2}\delta_1, D_A=D_0 + \delta_D, D_{B,C} =  D_0 - \frac{1}{2}\delta_D, I_A=I_0 + \delta_I, I_{B,C} =  I_0 - \frac{1}{2}\delta_I$. The stability matrix ${\cal M}_U$, is then
\begin{equation}
{\cal M}_U = \left(\begin{array}{ccc} -k_0 D_0 - \gamma & (-k_c+k_t/2)N_0 & \frac{d\tilde{\lambda}_N}{dI_0}  \\
(-k_c+k_t/2)D_0 & -k_0 N_0 - \gamma &  \frac{d\tilde{\lambda}_D}{dI_0}  \\
k_t \lambda_I D_0 & -(k_t/2)\lambda_I N_0 & -\gamma_I \end{array}\right)
\end{equation}
The instability sets in when $\textrm{Det} {\cal M}_U = 0$, i.e., when
\begin{equation}
k_t\lambda_I D_0 \left[4\gamma + 3 (k_0 + k_c)N_0\right]\frac{d\tilde{\lambda}_N}{dI_0} =
4\gamma_I \gamma \left[\gamma + k_0\left(D_0+N_0\right)\right] + 3\gamma_IN_0D_0k_t\left(k_0+3k_c\right) + 2\frac{d\tilde{\lambda}_D}{dI_0}k_t\lambda_IN_0\left(\gamma + 3k_cD_0\right)
\label{eq:stab}
\end{equation}

To derive the condition for the bifurcation point to be a pitchfork, we consider a general nonlinear system of steady-state equations, $f_i({x_j},\mu)=0$, where the functions $f_i$ depend on the variables ${x_j}$ and the bifurcation parameter $\mu$.  The steady-state at the bifurcation point, $\mu_0$ is denoted $x_{j,0}$. At a slightly shifted value of $\mu=\mu_0 + \Delta\mu$, the solution is given by $x_j = x_{j,0} + \epsilon x_{j,1}$.  Expanding the steady-state equation to first order in $\mu$ and third order in $\\epsilon$ yields
\begin{equation}
0=\epsilon \frac{\partial f_i}{\partial x_j} x_{j,1} + \frac{\epsilon^2}{2} \frac{\partial^2 f_i}{\partial x_j\partial x_k}x_{j,1}x_{k,1} + \Delta\mu \frac{\partial f_i}{\partial\mu} +
\frac{\epsilon^3}{6}\frac{\partial^3 f_i}{\partial x_j\partial x_k\partial x_l} x_{j,1}x_{k,1}x_{l,1} + \epsilon\Delta\mu \frac{\partial^2 f_i}{\partial x_j\partial \mu}x_{j,1}
\end{equation}
where repeated indices are summed over and all derivatives are evaluated at the unperturbed solution. At the critical point, $x_{i,1}$ is given by the right zero-mode eigenvector of the linear stability operator  $\frac{\partial f_i}{\partial x_j}$, $x_{j,1} = x^R_j$.  At a pitchfork bifurcation, the second derivative terms have no projection on the zero-mode, so that
\begin{equation}
\frac{\partial^2 f_i}{\partial x_j\partial x_k}x^L_i x^R_{j}x^R_{k} = 0.
\end{equation}
The right and left eigenvectors are given by
\begin{equation}
x^R = \left(\begin{array}{c} 4\gamma_I (\gamma + k_0N_0) + 2k_t \lambda_I N_0 \frac{d\tilde{\lambda}_D}{dI_0} \\
-4\gamma_I\left(k_c - \frac{k_t}{2}\right) D_0 + 4k_t\lambda_I D_0 \frac{d\tilde{\lambda}_D}{dI_0} \\
k_t \lambda_I D_0\left(4\gamma + 6\left(k_c +  \frac{k_t}{2}\right)N_0\right) \end{array}\right); \qquad 
x^L = \left(\begin{array}{c} k_t \lambda_I D_0\left(4\gamma + 6\left(k_c +  \frac{k_t}{2}\right)N_0\right) \\-2k_t\lambda_IN_0(\gamma + 3k_cD_0)\\
4\gamma^2 + 4k_0\gamma( D_0+N_0) + 3k_t (k_0+3k_c)D_0N_0 \end{array}\right)
\end{equation}
Substituting this, the condition for a pitchfork bifurcation reads
\begin{align}
0=&\frac{d^2\tilde{\lambda}_N}{dI_0^2}k_t^3\lambda_I^3D_0^3\left[4\gamma+3(k_0+k_c)N_0\right]^3-2\frac{d^2\tilde{\lambda}_D}{dI_0^2}k_t^3\lambda_I^3D_0^2N_0(\gamma+3k_cD_0)\left[4\gamma+3(k_0+k_c)N_0\right]^2 \nonumber\\
&-8k_t \lambda_ID_0(\gamma +3k_cD_0)(2\gamma+3k_tN_0)\left[2k_t\lambda_I\frac{d\tilde{\lambda}_D}{dI_0}-\gamma_I(2k_c-k_t)\right]
\left[k_t\lambda_IN_0\frac{d\tilde{\lambda}_D}{dI_0}+2\gamma_I(\gamma+k_0N_0)\right]
\end{align}

For completeness, we note that the condition for a saddle-node bifurcation is in general
\begin{equation}
\frac{\partial^3 f_i}{\partial x_j\partial x_k\partial x_l} x^L_i x^R_{j}x^R_{k}x^R_{l} \frac{\partial f_m}{\partial\mu}x^L_m = 3\frac{\partial^2 f_i}{\partial x_j\partial x_k}x^L_i x^R_{j}x^R_{k}\frac{\partial^2 f_m}{\partial x_j\partial \mu}x^L_{m}x^R_n.
\end{equation}
Using the expressions for the eigenvectors above, one can use this to derive a fairly lengthy expression for the saddle-node condition.

\section{The Large $\lambda_D$ Limit and the Region of Hexagonal Instability} \label{sec:LargeLambdaD}
\setcounter{figure}{0} 
We showed in the main text the region of hexagonal instability in the $k_c$, $\lambda_D$ plane for various $\lambda_N$.  One striking feature of this diagram is that the phase boundaries become vertical at two critical values of $k_c$, so that $\lambda_D^c(k_c)$ diverges at these values.  We can derive this analytically by solve the steady-state equations in the large $\lambda_D$ limit.  

\begin{figure} 
\includegraphics[width=0.7\textwidth]{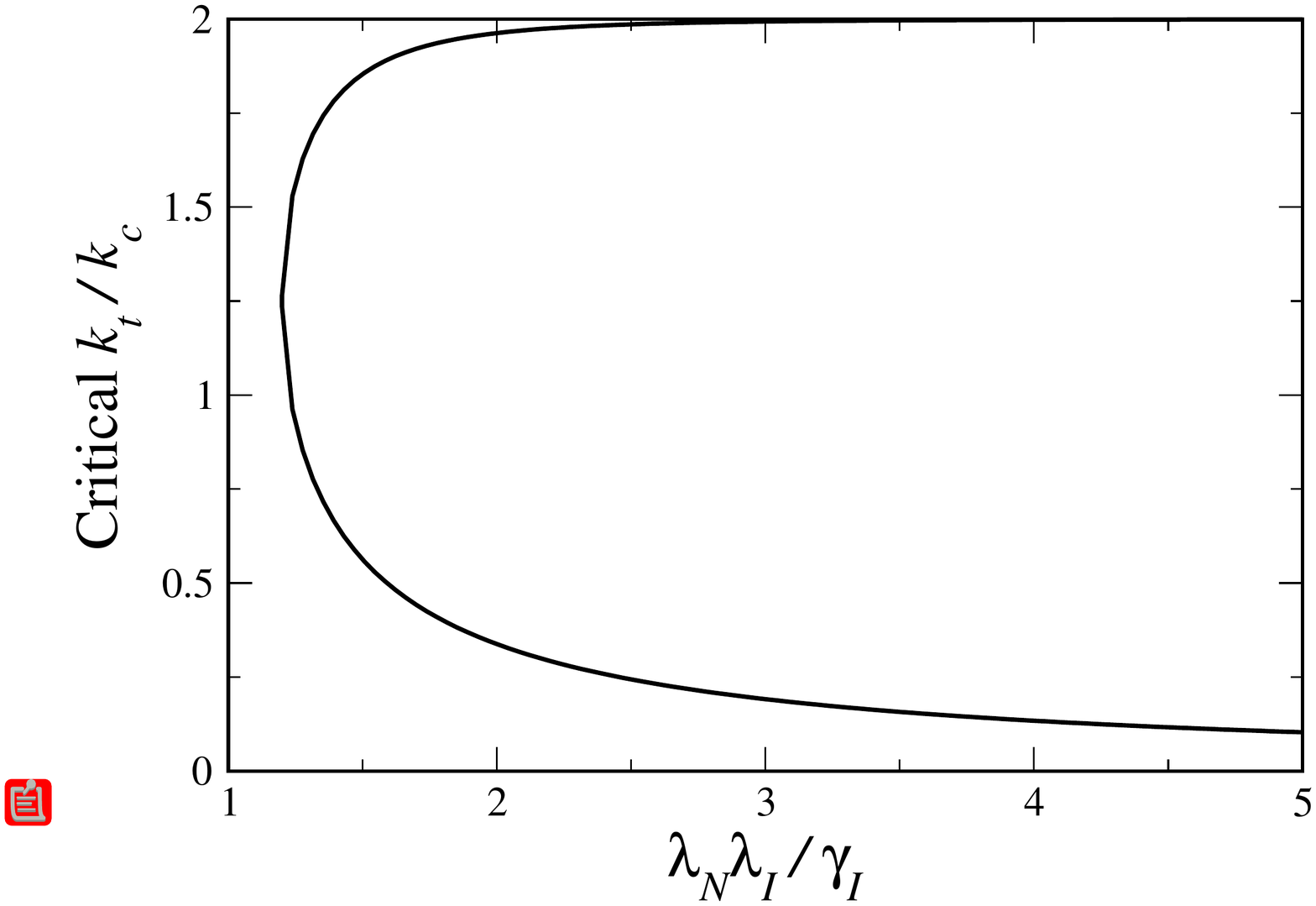}
\caption{The two critical values of $k_t/k_c$ as a function of $\xi=\lambda_N\lambda_I/\gamma_I$, for $k_H=1$ in the large $\lambda_D$ limit. We see that in line with our analytic calculation, for large $xi$, the critical $k_t/k_c$ either approaches 2 or 0.}
\label{figBigLamDkt}
\end{figure}

In this limit, $D_0\ll 1$ and $N_0 \ll 1$, we have
\begin{equation}
N_0D_0 = \frac{\tilde{\lambda}_N}{k_0} = \frac{\gamma_I I_0}{k_t\lambda_I}
\label{eq:I0}
\end{equation}
finite, which provides a closed equation for $I_0$. Then, the stability condition Eq. \eqref{eq:stab} reduces to
\begin{equation}
 \frac{d\ln \tilde{\lambda}_N}{d\ln I_0} = 1 + \frac{3k_c}{2k_0} \frac{d\ln\tilde{\lambda}_D}{d\ln I_0}
\end{equation}
Doing the algebra, we find that the critical $I_0$ satisfies the equation
\begin{equation}
3\gamma_I (1+I_0^2) I_0^3 = \lambda_I \lambda_N \left[ I_0^2(1+2I_0^2)(1+k_H)-1\right]
\end{equation}
This equation typically has two positive roots for $I_0$, which when substituted back into Eq. \eqref{eqI0} yields the critical value of $k_0/k_c=1 + k_t/k_c$. We also see that the critical $k_t/k_c$ only depends on the lumped parameter $\xi\equiv\lambda_I \lambda_N/\gamma_I$ and $k_H$. For large $\xi$, things simplify further and we find the solutions
\begin{equation}
I_0 \approx \frac{2}{3}\xi (1+k_H); \qquad\qquad  I_0 \approx \frac{1}{2}\left[\sqrt{\frac{k_H+9}{k_H+1}}-1\right]^{1/2},
\end{equation}
corresponding to 
\begin{equation}
\frac{k_t}{k_c}\approx 2; \qquad  \frac{k_t}{k_c} \approx \frac{1+k_H}{\xi} \left[\sqrt{\frac{k_H+9}{k_H+1}}-1\right]^{1/2}.
\end{equation}
In Fig. \ref{figBigLamDkt}, we plot the two solutions for $k_t/k_c$  as a function of $\lambda_N$ for the standard parameters $k_H=1$, $\lambda_I=1$, $\gamma_I=1/2$.  We see that for $\xi\gtrsim 1$, there are two critical values of $k_t/k_c$, one near 0 and the other near 2, between which the uniform solution is unstable to a hexagonal pattern.  This behavior is manifest in the numerical unstable region plotted in Fig. \ref{figstability} in the main text.

\section{The Large $\lambda_N$ Limit and the Region of Hexagonal Instability} \label{sec:LargeLambdaN}
\setcounter{figure}{0}
Similarly, we can compute the region of hexagonal instability for large $\lambda_N$.  Here, $N_0$ is large and $D_0$ is small, with the product being of order 1: \begin{equation}
N_0 D_0 \approx  \frac{\tilde{\lambda}_D}{k_0} = \frac{I_0\gamma_I}{k_t \lambda_I}
\label{eq1}
\end{equation}
 Then, the stability criterion reads
 \begin{equation}
 \frac{d\ln \tilde{\lambda}_D}{d\ln I_0} = \frac{3(k_c + k_0)}{2k_0}  \frac{d\ln \tilde{\lambda}_N}{d\ln I_0}  - 2
 \end{equation}
 which reduces to
 \begin{equation}
1 = \frac{I_0^2}{1+I_0^2} + \frac{3I_0^2k_H(k_c + k_0)}{2k_0(1+I_0^2)(1+I_0^2(1+k_H))}.
\end{equation}
Eliminating $k_t/k_c$ yields an equation for $I_0$ in terms of $k_H$ and the lumped parameter $\zeta=\lambda_D \lambda_I/\gamma_I$
\begin{equation}
3I_0^3 k_H = 2 \left[1 + I_0^2(1+2k_H)\right ]\zeta
\end{equation}
yielding for $I_0$,
\begin{equation}
I_0^2 = \frac{2k_0}{k_c(4k_H-2) + k_t(k_H-2)} . 
\end{equation}
Substituting this into Eq. \eqref{eq1} and solving for $\zeta$ yields
\begin{equation}
\zeta=\sqrt{2} k_H \frac{k_t+4k_c}{k_t} \left(\frac{k_t+k_c}{ k_c(4k_H-2) + k_t(k_H-2)}\right)^{3/2}
\end{equation}
This solution, which specifies the critical value of $\zeta$ (or $\lambda_D$ for given $\lambda_I$ and $\gamma_I$) is physical only if $k_c/k_t > (2-k_H)/2(2k_H-1)$. 
This relation, plotted in Fig. \ref{figstability} for $k_H=2/3$ and $k_H=4/3$ shows that $\zeta$ decreases from infinty as $k_t$ increases from 0, reaches a minimum and then increases, diverging as $k_t/k_c$  approaches $(2-k_H)/2(2k_H-1)$.  The graph is similar for all  $1/2\le k_H < 2$. For $k_H<1/2$ there is no critical $\lambda_D$ for large $\lambda_N$, whereas for $k_H>2$, there is a critical $\lambda_D$ for all $k_t/k_c$.  

It should be noted that our results for large $\lambda_D$ indicated that that the point of divergence of $\lambda_D$ is close to $k_t = 2k_c$ for large $\lambda_N$.
However, our current large $\lambda_N$ calculation indicates that $\lambda_D$ diverges at $k_t=[(2-k_H)/2(2k_H-1)]k_c$.  This indicates that the large $\lambda_D$ and large $\lambda_N$ limits do not commute.  This is reflected especially in the right panel of Fig. \ref{figstability} in the main text, where for $k_H=4/3$, we see that for large finite $\lambda_N$, $\lambda_D$ appears to be diverging at the location predicted by the large $\lambda_N$ calculation, but once $\lambda_D$ is sufficiently large, it turns back at the curve eventually approaches the $k_c=2k_t$ line.

\section{Multiple Uniform States} \label{sec:multiple}
\setcounter{figure}{0}  We have found through numerical experiments parameter regions that show the existence of multiple uniform states.  Here we want to clarify where in parameter space such solutions exist. We start by again assuming large
 $\lambda_D$. We get, following the same scaling as above, 
 \begin{equation}
 I_0 \ = \ {\cal{R}}(I_0) \equiv  \frac{ k_t \lambda_N(1 + k_H \frac{I_0^2}{1+I_0^2}) }{\gamma_I k_0} . \label{eq0A}
 \end{equation}
There are either 1 or 3 solutions of this nonlinear equation. The bifurcation point is where these three solutions collapse to 1.  At this point,
 \begin{equation}
 0 =  {\cal{R}}' - 1 =  {\cal{R}}'' .
 \end{equation}
 Denoting $Q\equiv  \frac{ k_t \lambda_N}{\gamma_I k_0}$, this reads
 \begin{align}
 1 &= 2Qk_H \frac{I_0}{(1+I_0^2)^2}, \nonumber\\
 0 &= 2Q k_H \frac{1-3 I_0^2}{1 + I_0^2)^3}.
 \end{align}
 Thus, from the second of these, at the bifurcation, $I_0^* = \sqrt{3}/3$. The first then implies $k_H^* = \frac{8\sqrt{3}}{9 Q*}$. Plugging these into
 Eq. \ref{eq0A}  gives 
 \begin{equation}
 Q^*=\frac{\sqrt{3}}{9},
 \end{equation}
 so that
 \begin{equation}
 k_H^* = 8.
 \end{equation}
 To see how things look in the vicinity of the bifurcation point, we write $Q=Q^* + \delta_Q$, $k_H = k_H^* + \delta_k$, $I_0 = I_0^* + \delta_I$. Expanding to third order in $\delta_I$,  Eq. \eqref{eq0A} translates to
 \begin{equation}
 0 = 1/36 (\sqrt{3} \delta_k + 108 \delta_Q + 9 \delta_k \delta_Q) + 
 1/8 (\delta_k + 24 \sqrt{3} \delta_Q + 3 \sqrt{3} \delta_k\delta_Q) \delta_I - 
 3/32 (8 + \delta_k + 24 \sqrt{3} \delta_Q + 3 \sqrt{3} \delta_k \delta_Q) \delta_I^3.
 \end{equation}
 In the last term, we can drop all the $\delta_k$, $\delta_Q$ pieces as being higher order. Then, for the second term to be the same order as the third, we have to have that $\delta_k$, $\delta_Q$ are both of order $\delta_I^2$, and we can drop the last piece of the second term. To make the first term also of the same order, its leading contribution must vanish, so that $\delta_Q = - \sqrt{3}/108 \delta_k + \alpha \delta_k^{3/2}$. Then, the equation reduces to
 \begin{equation}
 0 = \frac{1}{36} \left( 108 \alpha \delta_k^{3/2}\right ) + \frac{1}{8} \left(\frac{2}{3} \delta_k\right) \delta_I - \frac{3}{4} \delta_I^3 .
 \end{equation}
To have three solutions, we must have $\delta_k>0$, so that the derivative of this equation has two solutions. Then, then are solutions in a band around $\delta_Q = - \sqrt{3}/108 \delta_k $, which is negative.  Performing the numerical solution for large but finite $\lambda_D$ yields the same conclusions, with $k_H^*$ rising as $\lambda_D$ falls. As we take $\lambda_D$ smaller and smaller, $k_H^*$ continues to rise, and eventually diverges at a finite value of $\lambda_D$. This is shown in Fig. \ref{fig:Unif}. We will calculate this value momentarily, but the implication is clear: there are only multiple uniform solutions for $k_H>8$, which is much larger than physical.  This is all the more so since $\lambda_D$ is finite, and the multiple uniform solution lower bound on $k_H$ is thus in fact much higher.
 \begin{figure}
 \begin{center}
 \includegraphics[width=0.8\textwidth]{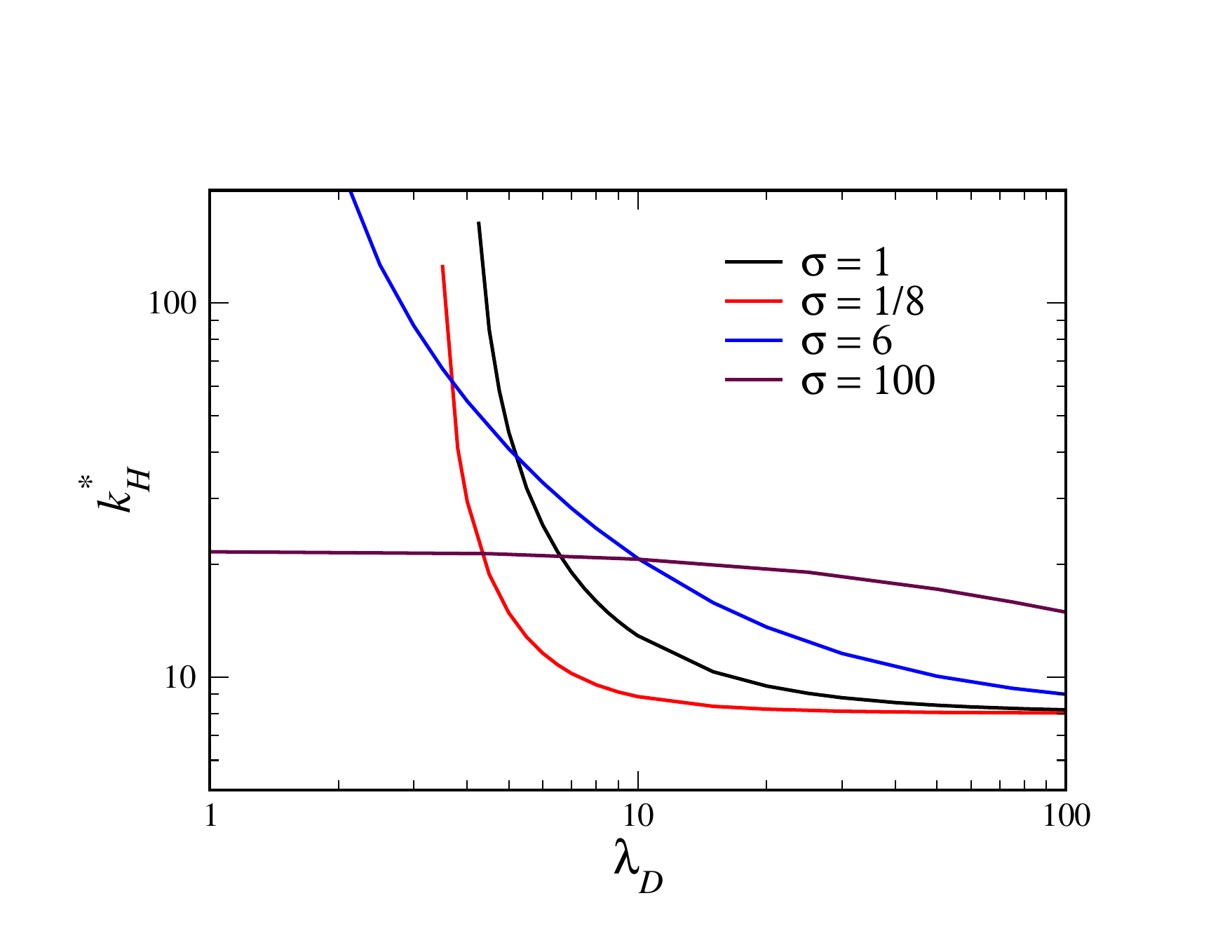}
 \end{center}
 \caption{The critical value $k_H^*$ above which multiple uniform solutions are possible (for some range of $Q$).  $\sigma \equiv \frac{\gamma_D\gamma_N}{k_0 \lambda_N}=1$, $1/8$, and $6$, and $100$.} 
 \label{fig:Unif}
 \end{figure}
 
Our last task is to determine the minimal value, $\lambda_D^*$ below which multiple uniform solutions are impossible. The numerics show that as $\lambda_D$ approaches its critical value, $I_0^*$ approaches 0 as $k_H^*$ diverges, with $k_H^*( I_0^*)^2$ remaining finite. In this limit, denoting $x=k_H I_0^2$, we have
\begin{equation}
R_0(I_0) = \frac{Q}{2} \left[\frac{\lambda_D}{\lambda_N} + \frac{\gamma_D\gamma_N}{k_0 \lambda_N} + 1 + x - \sqrt{\left(\frac{\lambda_D}{\lambda_N} + \frac{\gamma_D\gamma_N}{k_0 \lambda_N} + 1 + x \right)^2 - 4\frac{\lambda_D(1+x)}{k_0}}\right]
\end{equation}
We introduce the notation $\sigma=\frac{\gamma_D\gamma_N}{k_0 \lambda_N}$, $A=\frac{\lambda_D}{\lambda_N} + \sigma + 1$, $B=4\frac{\lambda_D}{\lambda_N}$ and rewrite
\begin{equation}
R_0(I_0) = \frac{Q}{2} \left(A + x - \sqrt{(A + x )^2 - B(1+x)}\right)
\end{equation}
At the bifurcation point, we have $R_0(I_0)=0$, $R_0'(I_0)=0$, and $R_0''(I_0)=0$, where the derivatives are not w.r.t. $x$. Proceeding, we find the solution
\begin{equation}
x^*=A/3;  \qquad\qquad B^*=\frac{4A^2}{3}(1 -  A/9);\qquad\qquad Q^* = \frac{3  \sqrt{3}}{\sqrt{A}( 9-A)}
 \end{equation}
Given $B^*$, since 
\begin{equation}
1+  \frac{\gamma_D\gamma_N}{k_0 \lambda_N} = A - B/4
\end{equation}
we have that
\begin{equation}
A = 3\left( 1 + \left[ \sigma\right]^{1/3}\right)
\end{equation}
This corresponds to a critical value of $\lambda_D/\lambda_N$ at which $k_H^*$ diverges of
\begin{equation}
\frac{\lambda_D}{\lambda_N} =  \left[ 1 +\sigma^{1/3}\right]^2 \cdot \left[2 - \sigma^{1/3}\right]
\end{equation}
Examining this, we see that there is a maximum of $\lambda_D/\lambda_N$ for $\sigma=1$.  Also, 
 there is a change of behavior for $\sigma>8$. Above this value, $k_H^*$ remains finite down to $\lambda_D=0$, and $\lambda_D=0$ it is a decreasing function of $
 \sigma$, approaching $10+4\sqrt{6}$ from above as $\sigma \to\infty$,   In general, as $\sigma$ gets large,
 $k_H^*$ is a monotonically decreasing function of $\lambda_D/(\lambda_N\sigma)$, going between $10+4\sqrt{6}$ at $\lambda_D/\sigma\to0$ and 8 as $\lambda_D/\sigma\to\infty$:
 \begin{equation}
k_H^*= \frac{320 + 464 \frac{\lambda_D}{\lambda_N\sigma} + 160 \left( \frac{\lambda_D}{\lambda_N\sigma}\right)^2 + 7\left( \frac{\lambda_D}{\lambda_N\sigma}\right)^3 + \sqrt{96 + 96 \frac{\lambda_D}{\lambda_N\sigma} + 9 \left(\frac{\lambda_D}{\lambda_N\sigma}\right)^2}\left(32 + 32 \frac{\lambda_D}{\lambda_N\sigma} + 3\left( \frac{\lambda_D}{\lambda_N\sigma}\right)^2\right)} { 2 \left(1 + \frac{\lambda_D}{\lambda_N\sigma}\right)\left(16 + 16 \frac{\lambda_D}{\lambda_N\sigma} + \left(\frac{\lambda_D}{\lambda_N\sigma}\right)^2\right)}
 \end{equation}

\section{Hexagonal Symmetry Steady-State Equation\label{sec:hex}}
\setcounter{figure}{0}
As discussed in the main text, one can reduce the full system of equations to a coupled ODE system. If we specialize to the case of steady-state patterns with hexagonal symmetry, so that the $B$ and $C$ sublattices are equivalent, we obtain the 6-dimensional system 
\begin{eqnarray}
0 & = &  \lambda_N H_+ (I_A)  -N_A \left (k_c D_A + k_t D_B    \right) - \gamma N_A \nonumber \\
0 & = &  \lambda _D  H_- (I_A)  -D_A \left (k_c N_A + k_t N_B  \right) - \gamma D_A\nonumber \\
0  & = &  k_t N_A D_B  -\gamma _I I_A \nonumber \\
0 & = &  \lambda_N H_+ (I_B)  -N_B \left (k_c D_B + \frac{k_t }{2} (D_A  +D_B)   \right) - \gamma N_B \nonumber \\
0 & = &  \lambda _D  H_- (I_B)  -D_B \left (k_c N_A + \frac{k_t }{2}(N_B  + N_A)  \right) - \gamma D_B\nonumber \\
0  & = &  \frac{k_t }{2}\lambda_I N_B (D_A+ D_B)  -\gamma _I I_A
\end{eqnarray}

\section{Hi-Hi Solutions\label{HH}}
\setcounter{figure}{0}
 In the main text, we discussed the fact that the model as presented can support anomalous solutions in which the levels of Notch and Delta are positively correlated in the different sub-lattice sites instead of being anti-correlated.  
 To study this in more detail, we look analytically at the large $\lambda_D$ limit and look for the critical curve along which Notch on the $A$ and $B$ sublattice sites are equal. This curve should demarcate the boundary between regular hexagonal solutions and what we are calling Hi-Hi solutions. Studying the numerics (data not shown), we see that $D_A$ and $D_B$ are large, of order $\lambda_D$ and $N_A=N_B\equiv N$ is small, of order $1/\lambda_D$. Writing $N_A=N_B=N_0/\lambda_D$, $D_A=\lambda_D D_{A0}$, $D_{B}=\lambda_D D_{B0}$ and specializing to the case of Hill coefficient 2 in the transcriptional/regulation terms, we have to solve the system
 \begin{align}
 \lambda_N \left(1 + \frac{k_H I_A^2}{1 + I_A^2}\right) &=k_T N_{0} D_{B0} + k_cN_0 D_{A0} ,
 \nonumber\\
  \lambda_N \left(1 + \frac{k_H I_B^2}{1 + I_B^2}\right) &=\frac{1}{2}k_T N_{0}(D_{A0}+ D_{B0}) + k_cN_0 D_{B0} ,\nonumber\\
 \frac{1}{1+I_A^2} &= \gamma D_{A0} ,\nonumber\\
 \frac{1}{1+I_B^2} &= \gamma D_{B0} ,\nonumber\\
 k_t \lambda_I N_0 D_{B0} &= \gamma_I I_A,\nonumber\\
 \frac{1}{2} k_t  \lambda_I N_0 (D_{A0}+D_{B0}) &= \gamma_I I_B.
 \end{align}
 Here we have chosen units such that $s_0=1$.  This system admits two exact solutions, in terms of $\eta \equiv \frac{\lambda_N \lambda_I (1+k_H)}{\gamma_I}$:
 \begin{align}
 I_A &=\frac{ \eta \pm \sqrt{ \eta^2 - 12}}{3} \nonumber\\
 I_B &= \frac {5 \eta \mp \sqrt{\eta^2-12}}{6} \nonumber\\
 D_{A0} &= \frac{1}{\gamma (1 + I_A^2)} \nonumber\\
 D_{B0} &= \frac{1}{\gamma (1 + I_B^2)} \nonumber\\
 N_0 &= \frac{\gamma \gamma_I I_A(1+I_B^2)}{ k_t } \nonumber\\
 k_c &= k_t \frac{\eta^2 (17 - k_H) \pm 3k_H \eta \sqrt{ \eta^2-12}  \pm
   2k_H \eta^3 \sqrt{ \eta^2-12}  + 4 (1 + k_H) + 
   2 \eta^4 (2 + k_H) }{2 (4 + \eta^2) (1 + 4 \eta^2) (1 + k_H)}
 \end{align}
 The two solutions merge at $\eta=\sqrt{12}$ and do not exist for $\eta<\sqrt{12}$. One key result is that the critical $k_c$ is proportional to $k_t$. For large $\eta$, which is typical for our parameters where $\gamma=0.1$, things simplify tremendously and we have for the ``$-$" branch:
  \begin{align}
 I_A &=\frac{ 2}{\eta}\ll 1 ,\nonumber\\
 I_B &= \eta \gg 1,\nonumber\\
 D_{A0} &= \frac{1}{\gamma}, \nonumber\\
 D_{B0} &= \frac{1}{\gamma \eta^2}  \ll 1,\nonumber\\
 N_0 &= \frac{2\gamma\gamma_I \eta}{ \lambda_Ik_t } \gg 1, \nonumber\\
 k_c &=  \frac{k_t}{2(1+k_H)}.
 \end{align}
 In fact, for $\eta=\sqrt{12}$ which is its lower limit, we get $k_c/k_t = (14+5k_H)/28/(1+k_H)$, which is only slightly higher than the infinite $\eta$ result.
 
 Going off this border line numerically, we find that the solution to the left of the line is a Hi-Hi solution, whereas to the right is a regular hexagon.
 The branch starting from the ``$-$" border is stable whereas starting from the ``$+$" border is unstable.  Thus, the two solutions are completely distinct, even in the region to the right of both borders. The region of stable Hi-Hi solutions extends all the way down to $k_c=0$. The relevant phase diagram is shown in Fig. \ref{lowkC}. We show the line above which stable Hi-Hi solutions exist, and below which stable Hi-Lo hexagons exist, both of which coexist with a stable uniform solution. We also show the saddle-node bifurcation line, below which only the uniform solution is stable.
 


\end{document}